\documentclass[11pt]{article}
\usepackage{times}
\usepackage{amsfonts}
\usepackage{amsmath}
\usepackage{enumitem}
\usepackage{mathtools}
\usepackage{centernot}
\usepackage{soul}
\usepackage{booktabs}
\usepackage{latexsym}
\usepackage[ruled]{algorithm2e}
\SetAlFnt{\small}
\SetAlCapFnt{\small}
\SetAlCapNameFnt{\small}
\SetAlCapHSkip{0pt}
\IncMargin{-\parindent}
\usepackage{algorithmic}
\usepackage{nicefrac}
\usepackage[
  bookmarks=true,
  bookmarksnumbered=true,
  bookmarksopen=true,
  pdfborder={0 0 0},
  breaklinks=true,
  colorlinks=true,
  linkcolor=black,
  citecolor=black,
  filecolor=black,
  urlcolor=black,
]{hyperref}
\usepackage{graphicx}
\usepackage{xcolor}
\usepackage[margin=1in]{geometry}
\usepackage{amssymb}
\usepackage{amsmath}
\usepackage[square]{natbib}
\setcitestyle{numbers}
\usepackage{amsthm}
\usepackage[capitalize]{cleveref}
\usepackage{bbm}
\usepackage{MnSymbol}

\Crefname{equation}{Equation}{Equations}

\theoremstyle{plain}
\newtheorem{theorem}{Theorem}

\newtheorem{fact}[theorem]{Fact}

\newtheorem{lemma}[theorem]{Lemma}
\newtheorem{proposition}[theorem]{Proposition}
\newtheorem{remark}[theorem]{Remark}
\theoremstyle{definition}

\newtheorem{notation}[theorem]{Notation}
\newtheorem{example}[theorem]{Example}
\DeclarePairedDelimiter\parentheses{(}{)}
\DeclarePairedDelimiter\braces{\{}{\}}
\DeclarePairedDelimiter\brackets{[}{]}
\DeclarePairedDelimiter\absolute{|}{|}

\begin{document}
\newcommand{\supp}{{\mathrm{supp}}}
\newcommand{\poly}{{\mathrm{poly}}}
\newcommand{\co}{{\mathrm{co}}}
\newcommand{\mc}{{\mathrm{MC}}}

\title{Designing Information in a Forest: Algorithmic Aspects}
\date{November 12, 2026}
\author{Konstantin Zabarnyi\thanks{Yale University| \emph{E-mail}: \href{mailto:konstzab@gmail.com}{konstzab@gmail.com}}}
\maketitle

\begin{abstract}
The Bayesian persuasion framework models strategic information revelation by a sender to one or more receivers. We study an intricate version of the multi-receiver persuasion setting in which the communication hierarchy is a \emph{forest} -- i.e., some receivers observe more informative signals than others, and the partial ordering of the receivers according to the information they get is represented by a pre-specified directed forest. This model naturally captures interactions in hierarchical organizations, and it lies in between the two most commonly studied versions of Bayesian persuasion: private persuasion and public persuasion. Inspired by the classic model of private persuasion, we focus on a setting with a binary state of nature and binary-action heterogeneous receivers s.t. the sender's utility is a monotone function of the set of receivers taking a specific desirable action.

We extend the computational tractability result for symmetric sender's utility from private persuasion to the forest hierarchy framework using methodology inspired by optimal transport theory. We further provide a polynomial-time algorithm when the sender's utility is modular. On the negative side, we show that, unlike in private persuasion, the case of a supermodular sender's utility function leads to an NP-hard optimization problem which further does not admit an FPTAS unless P=NP; this result holds even when each tree component of the forest is a chain of 3 receivers. To complement the hardness result, we provide a simple approximation algorithm with the approximation ratio depending linearly on the depth of the forest hierarchy. We further provide an analogous algorithm for the submodular utility case. Finally, beyond forest communication hierarchies, we rule out an FPTAS for the sender's optimal utility even in the symmetric utility case, unless P=NP.
\end{abstract}

\section{Introduction}
\label{sec:intro}
Information in large companies is often neither publicly available to all employees nor perfectly private. For example, a supplier may provide a report about an offered product to a certain department, yet such a report may be inspected by the branch head. While parallel departments have access to different information sources, shared supervision and reporting often lead to cross-cutting access to information. In this paper, we study optimal information disclosure policies when information observed by different agents adheres to a certain hierarchy of informativeness levels. We focus on providing algorithmic descriptions of optimal policies for a wide range of economically meaningful objectives.

\paragraph{A leading example.} Suppose that a supplier (\emph{sender}) aims to sell a certain product to multiple companies. The product has some set of characteristics (\emph{state}) which is not publicly known. Each company has managers (\emph{receivers}) in different tiers; every manager makes the procurement decisions for their own unit. The managers have some common-knowledge \emph{prior belief} regarding the characteristics of the product based, e.g., on empirical studies. The supplier can design an experiment for each manager that will reveal partial information about the product characteristics to that specific manager directly. However, all the supervisors of each manager will be able to observe the information revealed to that manager. The supervisory relations create a (rooted) \emph{tree information hierarchy} within each organization -- i.e., the chair observes all the available information within the organization, the head of every branch observes all the information available to lower-tier managers within the branch, but does not observe the information provided to members of parallel branches, etc. The presence of multiple companies naturally leads to an information access hierarchy represented by a \emph{directed forest}.

The utility of the supplier depends on the set of managers who decide to purchase the product. E.g., the supplier may want a certain minimum number of managers to purchase the product. Alternatively, if the managers are responsible for units that serve overlapping markets, then increasing the number of purchasing managers will often yield diminishing returns; in contrast, if the units serve complementary markets, increased adoption may yield increasing returns. Finally, if the aim is to maximize total sales contribution margin, then modular objective functions naturally arise (see, e.g.,~\cite{horngren2009cost}).

\paragraph{The model.} This paper explores how a given communication hierarchy affects the design and computational tractability of the sender's optimal information revelation policy. We consider a Bayesian persuasion setting with a well-informed sender who can send signals to multiple receivers subject to respecting a given hierarchy of the receivers in terms of the information they observe. As is standard, we call the sender's information revelation policy a \emph{signaling scheme}, and we focus on computing (approximately) optimal signaling schemes for the sender in the sense of maximizing the sender's utility. Similarly to the works of~\cite{babichenko2016computational,arieli2019private} that initiated the study of multi-receiver persuasion under the assumption that signaling is private, we assume that each receiver has two actions, which we treat as \emph{adopt} or \emph{reject} a certain offer. The sender's utility is a monotone function of the set of receivers who adopt. Similarly to the foundational works of~\cite{babichenko2016computational,arieli2019private}, we assume that the state is binary.\footnote{In the latest version of their paper,~\cite{arieli2019private} partially generalize their non-computational results to a more general setting.} Our setting generalizes the private persuasion setting. Moreover, by~\cite{babichenko2022multi}, the sender's utility in out setting is always upper-bounded by her utility in private persuasion.

The sender-receivers communication occurs through multiple channels that may overlap; the communication is specified by the given hierarchy among the receivers. We say that receiver $i$ \emph{(strictly) dominates} receiver $j$ if every signal observed by receiver $j$ must be observed also by receiver $i$, but the sender can transmit a signal to receiver $i$ without revealing any information to receiver $j$. As a simple example, if there are two receivers and two communication channels s.t.~receiver $1$ observes both channels but receiver $2$ observes only one of the channels, then receiver $1$ dominates receiver $2$. The partial order on the set of receivers specified by the domination relation imposes constraints on the ability of the sender to communicate with receivers. We refer to the directed graph specifying this partial order as the \emph{communication hierarchy}. 

\paragraph{The focus.} Throughout most of the paper, we focus on \emph{forest communication hierarchies}. Namely, we assume that the graph representing the domination relation is a rooted directed forest. Such structures naturally arise in hierarchical organizations. In particular, this class of hierarchies contains chains, which correspond to vertical information spillovers~\cite{mathevet2020organized}. Beyond the natural economic interpretation of forest communication hierarchies, they have an interesting property that will be key to our analysis: A hierarchy of receivers' beliefs according to information about the state in the Blackwell sense can always be implemented through refined signaling if the communication hierarchy of the receivers is a forest~\cite{brooks2022information}.

Motivated by the leading example, we are especially interested in supermodular, submodular, modular, or symmetric sender's utility function families. Given the computational tractability benchmarks for these classes set by~\cite{babichenko2016computational} for private persuasion, which is a special case of our model in which the forest has no edges, we ask the following question: \textbf{Can we exactly match the computational tractability results for forest communication hierarchies to the benchmarks provided by private persuasion?}

The following example shows that the hierarchy among receivers affects the optimal sender's utility in an intricate way.

\begin{example}
    \label{ex:illustrate} Consider a setting with a uniform prior distribution over the binary state space $\{0,1\}$ and $n=3$ receivers that have respective utility functions $v_1,v_2,v_3$ and actions $a_1,a_2,a_3\in\{0,1\}$. Suppose that the utility of each receiver for taking action $a_i=0$ is $0$ regardless of the state, while their utilities for picking $a_i=1$ depend on the state $\omega\in\{0,1\}$ as given in Table~\ref{tab:ut}. Assume further that the sender's utility is $1$ if at least two receivers take action $1$ and is $0$ otherwise. For $i=1,2,3$, let $p_i$ be the probability for the state to be $1$ conditional on the information that receiver $i$ gets from the sender (i.e., the \emph{posterior belief}). Under the standard assumption that the receivers are risk-neutral and break ties in the sender's favor, receiver $1$ prefers action $1$ iff $p_1\ge 1/2$, while the thresholds for receivers $2$ and $3$ are $p_2\ge 2/3$ and $p_3\ge 2/3$, respectively.

    \begin{itemize}
        \item Under private signaling, it is optimal to send a totally non-informative signal to receiver $1$ while anti-correlating the messages sent to receivers $2,3$ when $\omega=0$ (see Table~\ref{tab:sc}, left). This ensures that the sender gets expected utility of $1$, as always at least $2$ receivers pick $a_i=1$.
        \item If the receivers are arranged in a chain $1\implies 2\implies 3$ (i.e., receiver $1$ dominates both $2,3$ and receiver $2$ dominates $3$), we claim that the optimal expected sender's utility is $3/4$. Indeed, it is well-known that in Bayesian persuasion, the expected value of the posterior must equal the prior~\cite{blackwell1950comparison,aumann1995repeated,kamenica2011bayesian}. That is, $\mathbb{E}\brackets*{p_1}=\mathbb{E}\brackets*{p_2}=\mathbb{E}\brackets*{p_3}=1/2$.  Moreover, since receiver 2 observes all the information available to receiver 3, we can similarly deduce $\mathbb{E}\brackets*{p_2 | p_3}=p_3$; here, $p_3$ plays the role of an ``updated prior belief" for receiver 2 after observing the signal sent directly to receiver 3 and before observing the additional information available to him. Thus, Markov's inequality implies that $\Pr[p_3\ge 2/3]\le\frac{3}{4}$, and, moreover, for every fixed value of $p_3$, $\Pr[p_2\ge 2/3 | p_3]\le\frac{\mathbb{E}\brackets*{p_2 | p_3}}{2/3}=\frac{3 p_3}{2}$. Note that if at least two receivers take action $1$, then either $p_2\ge 2/3$ or $p_3\ge 2/3$. If $p_3< 2/3$, then we know that conditional on the value of $p_3$, $p_2\ge 2/3$ w.p.~at most $\frac{3 p_3}{2}$. Moreover, if $p_3\ge 2/3$, then conditional on the value of $p_3$, at least two receivers take action $1$ w.p. at most $1\le\frac{3 p_3}{2}$. Thus, the expected sender's utility is indeed upper bounded by $\mathbb{E}_{p_3}\brackets*{\frac{3 p_3}{2}}=\frac{3}{4}$, with equality achieved by sending the same message $m$ to all three receivers, where $m$ follows the conditional distributions of $m_3$ on $\omega$ from Table~\ref{tab:sc} (left).
        \item Suppose now that receiver $1$ dominates receivers $2,3$, while neither of receivers $2,3$ dominate the other one. Then trying to utilize the private signaling from Table~\ref{tab:sc} (left) will reveal the true state to receiver $1$. Thus, when $\omega=0$, exactly one receiver will take action $1$ (either receiver 2 or receiver 3), while when $\omega=1$, all three receivers will take action $1$, resulting in expected sender's utility of $1/2$. Trying to use the signaling scheme from the previous bullet will yield exactly the same expected sender's utility as before, namely $3/4$. It turns out that the expected sender's utility can be improved to $5/6$ (see Table~\ref{tab:sc}, right). It will follow from the algorithm provided by our Theorem~\ref{thm:chain} that this value is optimal for the given hierarchy of receivers.
    \end{itemize}
\end{example}

\begin{table}[h]
\begin{center}
\begin{tabular}{|c|c|c|}
\hline
      & $\omega=0$ & $\omega=1$ \\ \hline
$v_1$ & $-1$     & $1$    \\ \hline
$v_2$ & $-2$    & $1$    \\ \hline
$v_3$ & $-2$    & $1$ \\ \hline
\end{tabular}
\end{center}
\caption{The utility of each receiver $i=1,2,3$ for picking action $a_i=1$ as a function of the state $\omega$.}\label{tab:ut}
\end{table}

\begin{table}[h]
\begin{center}
\begin{tabular}{|c|c|c|}
\hline
 $\parentheses*{m_1,m_2,m_3}$     & $\omega=0$  & $\omega=1$ \\ \hline
$\parentheses*{1/2,2/3,0}$ & $1/2$     & $0$    \\ \hline
$\parentheses*{1/2,0,2/3}$ & $1/2$    & $0$    \\ \hline
$\parentheses*{1/2,2/3,2/3}$ & $0$    & $1$    \\ \hline
\end{tabular}
\hspace{5mm}
\begin{tabular}{|c|c|c|}
\hline
 $\parentheses*{m_1,m_2,m_3}$     & $\omega=0$  & $\omega=1$ \\ \hline
$\parentheses*{1/2,1/2,2/3}$ & $1/3$     & $1/3$    \\ \hline
$\parentheses*{1/2,2/3,1/2}$ & $1/3$    & $1/3$    \\ \hline
$\parentheses*{1,2/3,2/3}$ & $0$    & $1/3$    \\ \hline
$\parentheses*{0,0,0}$ & $1/3$    & $0$    \\ \hline
\end{tabular}
\end{center}
\caption{Optimal disclosure policies under private signaling (left) and signaling in a hierarchy s.t.~receiver $1$ dominates the other two receivers, who do not dominate each other (right); $m_i$ is the message sent directly to receiver $i$, and in this case it exactly equals $p_i$; each cell contains the probability to send the corresponding triple $\parentheses*{m_1,m_2,m_3}$ conditional on the state $\omega$.}\label{tab:sc}
\end{table}

\paragraph{Overview of results.} Our first main result is negative -- we show that unless $P=NP$, there is no fully polynomial-time approximation scheme (FPTAS) for approximately solving the sender's optimization problem when the sender's utility is supermodular, even if each component of the forest hierarchy of the receivers is a chain of just $3$ receivers (Theorem~\ref{thm:super_hard}). This contrasts with the private persuasion setting in which the supermodular utility case is computationally tractable~\cite{arieli2019private}. The focus on low-depth forests has a practical meaning. Indeed, in our leading example, it is natural to assume that the depth of the forest hierarchy is not too large. Fortunately, we derive a constant-factor approximation algorithm for the case of a supermodular sender's utility and an $O(1)$-depth forest hierarchy (Proposition~\ref{pro:super}). We further provide a constant-factor approximation algorithm for the optimal signaling scheme when the sender's utility is submodular and the forest has an $O(1)$ depth (Proposition~\ref{pro:sub}).

Our second main result is positive -- when the sender's utility is symmetric (i.e., it depends only on the number of receivers taking a certain action), then there is a polynomial-time algorithm computing an exactly optimal signaling scheme (Theorem~\ref{thm:chain} and Remark~\ref{rem:sym}). This is a generalization of a result of~\cite{babichenko2016computational}, but alternative proof methods are required. In particular, we only provide a succinct representation of an optimal signaling scheme, and the signaling scheme might require an exponential number of signals. To the best of our knowledge, this is the first computational tractability result for forest hierarchies with non-modular objectives. We also provide an algorithm for computing an optimal signaling scheme when the sender's utility is modular (Proposition~\ref{pro:linear}), which is an improvement of the additive FPTAS provided by~\cite{babichenko2022multi} (with the difference that we assume that the number of states is $2$ rather than $O\parentheses*{1}$ and the receivers are binary-action).

Finally, we show that when the hierarchy of the receivers is allowed to be a directed acyclic graph which is not a forest, then there is no FPTAS for approximating the sender's utility even when the sender's utility is a symmetric function taking only $3$ possible values and the hierarchy of the receivers is a directed bipartite graph, unless $P=NP$ (Proposition~\ref{thm:nonforest1}). This result has an important takeaway -- \emph{the forest property of the receivers' hierarchy has a crucial effect on efficient implementability of optimal information disclosure.} This takeaway can be seen as complementary to the findings of~\cite{brooks2022information} on the role of forest structures in constructability of the hierarchy of the receivers' beliefs through refinement of signals.

This result is also tightly connected to an analogous computational hardness result in a closely related setting of~\cite{candogan2020information}. Specifically,~\cite{candogan2020information} also prove that under hierarchical information access, if the hierarchy is not a forest, then the sender's optimization problem is, in general, NP-hard (and, unlike us, they show that it might hold even when the in-degree is at most $2$). However, our Proposition~\ref{thm:nonforest1} provides strengthening in the following sense: We assume that the number of states is just $2$ (rather than infinitely many); the sender's utility is monotone, symmetric, and takes only $3$ possible values; and the depth of the receivers' communication hierarchy (i.e., the maximum number of edges in a directed path) is $1$ rather than a variable number that can be arbitrarily large; none of these five conditions are satisfied by the construction of~\cite{candogan2020information}, and their result does not rule out an FPTAS. Therefore, our proof purifies the role of the receivers' hierarchy being a forest in computational tractability even when all the problem parameters are extremely small, and it shows that the hardness does not come from trade-offs in the sender's optimization problem caused by the non-monotonicity of her utility function. We should also mention that the result of~\cite{candogan2020information} on tractability of the non-forest hierarchy under a regime of information spillovers in which a message sent directly to some receiver only leaks to his immediate descendants does not undermine the crucial role forests play in tractability of persuasion under hierarchical information access: Indeed, in such a spillover regime, the presence of multiple information sources break dominating pairs of receivers, which significantly simplifies the directed graph representing the domination relation.

\paragraph{Our methodology.} The presence of intricate forest communication hierarchies requires using sophisticated methods. A standard approach in Bayesian persuasion is based on restricting attention to \emph{straightforward} signaling schemes -- i.e., each signal is treated as a recommendation of an action, and the sender designs the signaling scheme s.t.~it is always a best-response strategy for each receiver to follow the recommendation. This is the key tool which allowed~\cite{babichenko2016computational,arieli2019private} to significantly restrict the set of relevant signaling schemes and use common tools for optimizing set functions to derive their main results. A similar approach was used by~\cite{mathevet2020organized} to analyze a model in which the information is transmitted vertically in a chain. Their model differs from ours both because they focus on chain hierarchies and because they allow each receiver in the hierarchy to strategically choose whether to pass the signal to the next receiver in the hierarchy or not. A related approach based on nesting signals was also used by~\cite{mahzoon2022hierarchical}. While their methods do not directly translate to our setting, their idea of replacing the requirement that each action recommendation directly sent to some receivers is persuasive by requiring that the vector of recommendations observed by each receiver is persuasive inspired our key Lemma~\ref{lem:lp}, which characterizes the optimal signaling scheme through a linear program (LP). Lemma~\ref{lem:lp} is used in the proofs of both our main results. This lemma can be of independent interest, since \emph{it provides a full characterization of all inducible distributions over the receivers' joint beliefs conditional on the state} (subject to a certain discretization which can be done w.l.o.g.). Both the number of variables and the number of constraints in this LP are exponential even for forests with an $O(1)$ depth, and thus applying it to analyze computational aspects of our setting is far from being~immediate.

The second inspiration for the key Lemma~\ref{lem:lp} was the characterization of private Bayesian persuasion through optimal transport theory by~\cite{arieli2022persuasion}. While we do not use optimal transport terminology, Lemma~\ref{lem:lp} is a generalization of the characterization of~\cite{arieli2022persuasion} for feasible couplings of marginal distributions over the information of different receivers conditional on each state. However, we face significant additional challenges due to the forest hierarchy. Indeed, in private persuasion, there are no constraints on the coupling of additional information about the state revealed to different receivers \emph{in the same state}. Therefore, the challenge in coupling the marginal distributions arises only due to constraints connecting the information revealed in different states. In forest hierarchies, there are additional intricate constraints since some receivers observe signals transmitted to other receivers. To handle them, we combine the coupling idea of~\cite{arieli2022persuasion} with the nesting idea of~\cite{mathevet2020organized}.

Besides the practical justifications for focusing on forest hierarchies, there is also a technical reason: By~\cite{brooks2022information}, when a multi-channel communication structure is a directed forest, the Blackwell hierarchy of the receivers' beliefs can be implemented through nested signaling. This property seems to be crucial, since Proposition~\ref{thm:nonforest1} shows that even for very simple non-forest hierarchies (directed bipartite graphs), finding an optimal signaling scheme is NP-hard even when the sender's utility is symmetric and can take only $3$ different values.

We also note that the discretization technique used by~\cite{babichenko2022multi} to provide an additive FPTAS for forest hierarchies and modular sender's utilities cannot be applied to most parts of our paper since apart from Proposition~\ref{pro:linear}, we focus on non-modular objectives, for which the joint distribution over the beliefs of the receivers is crucial.

\subsection{Related Work}
\label{sub:related}

Since the introduction of the celebrated Bayesian persuasion model by~\citet{kamenica2011bayesian}, many of its extensions have been studied through the lens of economics~\cite{kamenica2019bayesian,bergemann2019information}, computer science~\cite{dughmi2017algorithmic}, and operations research~\cite{Candogan20}. Applications of Bayesian persuasion include advertising and marketing (see, e.g.,~\cite{kamenica2011bayesian,badanidiyuru2018targeting,drakopoulos2021persuading,candogan2022persuasion,babichenko2022regret,shin2024role}), recommender systems~\cite{che2018recommender}, voting~\cite{alonso2016persuading,castiglioni2020persuading,castiglioni2021persuading}, security games~\cite{xu2015exploring,rabinovich2015information}, revenue management~\cite{lingenbrink2018signaling,kuccukgul2022engineering}, and many more.

Most Bayesian persuasion literature focuses either on private persuasion (i.e., each receiver observes a separate signal privately)~\cite{bergemann2016bayes,alonso2016persuading,babichenko2016computational,arieli2019private} or on public persuasion (all receivers get exactly the same information)~\cite{bro2012send,dughmi2014hardness,cheng2015mixture,xu2020tractability,mathevet2020information}; some works compare both settings~\cite{wang2013bayesian,dughmi2017algorithmic}.

Our work adopts the classic model of private persuasion introduced by~\cite{babichenko2016computational} and studies whether analogous computational results hold for forest communication hierarchies; such hierarchies were first studied from the computational perspective by~\cite{candogan2020information,babichenko2022multi}. As discussed before, the key techniques from these papers (straightforward signaling schemes and using a discretization followed by a natural linear program) do not directly apply to prove our results due to our focus on non-modular objectives. The computational results on private persuasion by~\cite{babichenko2016computational,arieli2019private} serve as a benchmark for our setting.

A paper closely related to ours is~\cite{candogan2020information}. While both papers consider information spillovers among receivers with hierarchical information access,~\cite{candogan2020information} studies modular sender's objectives and a restricted class of partition-based signaling schemes for a continuum number of states and a variety of spillover regimes. Our warm-up Proposition~\ref{pro:linear} resembles the positive algorithmic result of~\cite{candogan2020information} for modular objectives and tree hierarchies, yet the assumptions in both settings are different -- while we focus on binary states and perfect information spillovers along the directed paths in the given receivers' hierarchy, we do not restrict the signaling schemes to a special class.

Among other works that study more intricate communication hierarchies than just private or public persuasion, we should mention~\cite{brooks2022information}, who focus on forest hierarchies;~\cite{galperti2019belief,galperti2020belief,kerman2025persuasion}, who study persuasion in networks; and~\cite{mathevet2020organized,mahzoon2022hierarchical}, who study persuasion with chain hierarchies. None of these papers focuses on computational~aspects. However, we rely on the main result of~\cite{brooks2022information} (Fact~\ref{fact:inducible}) multiple times in the paper.~\cite{castiglioni2021persuading} provide multi-criteria approximation algorithms for objectives representing voting in district-based elections and semi-public signaling -- i.e., the receivers are partitioned into disjoint subsets each of which gets a separate signal. Unlike our paper, the receivers in their framework do not have a tiered information access, and the communication channels do not overlap.~\cite{haghtalab2024leakage} study robustness of the sender's signaling scheme in private signaling to possible leakages of the private messages, under a leakage structure generated either adversarially or stochastically following the sender's choice of the signaling scheme. Unlike them, we focus on the sender's optimization problem under a given hierarchical information access.

Our linear programming (LP) characterization of the sender's optimization problem was inspired by optimal transport theory, which has been applied to Bayesian persuasion multiple times~\cite{arieli2022persuasion,dworczak2024persuasion,kolotilin2025persuasion,du2025optimal}. Our application is non-standard since the receivers may observe signals of some other receivers. A complementary approach to optimal transport (which aims to couple marginal distributions over receivers' beliefs conditional on the state) is provided by~\cite{arieli2021feasible}. That paper characterizes feasible distributions over the joint receivers' beliefs unconditional on the state, deriving a quantitative analogue of Aumann’s agreement theorem~\cite{aumann2016agreeing}. In contrast to the coupling conditional on the state, we were unable to apply the ideas of~\cite{arieli2021feasible} to forest hierarchies. The main obstacle is that it is unclear how to formulate the information hierarchy constraints in the joint unconditional distribution.

In our main negative result (Theorem~\ref{thm:super_hard}), we rule out the existence of an FPTAS (unless P=NP) when the sender's utility is supermodular, even when each tree component of the forest is a chain of just three receivers. To this end, we use a lexicographic reduction (see, e.g.,~\cite{behringer1977lexicographic,sherali1982equivalent,isermann1982linear}) from the max-cut problem. Namely, given a max-cut instance, we carefully construct a Bayesian persuasion instance with a forest communication hierarchy s.t.~the sender's utility is a sum of three components; every optimal signaling scheme must: (i)~maximize the expectation of the first component; (ii)~maximize the expectation of the second component subject to maximizing the expectation of the first component; and (iii)~maximize the expectation of the third component subject to satisfying properties (i) and (ii). A similar approach was used by~\cite{dughmi2016algorithmic} in a single-receiver Bayesian persuasion setting with action payoffs being independently but not identically distributed. However,~\cite{dughmi2016algorithmic} prove $\#$P-hardness rather than NP-hardness and the instance constructed in the reduction is substantially different.

\subsection{Paper Outline}
In Section~\ref{sec:prelim}, we introduce a model and present the preliminary Lemma~\ref{lem:grid} that allows us to discretize the sender's optimization problem. In Section~\ref{sec:mod}, we solve the case of a modular sender's utility in polynomial time. In Section~\ref{sec:supersub}, we present the key Lemma~\ref{lem:lp} and analyze tractability of the sender's optimization problem for a supermodular (Subsections~\ref{sub:super} and~\ref{sub:super_easy}) and a submodular (Subsection~\ref{sub:sub}) sender's utility. In Section~\ref{sec:sym}, we prove that the sender's optimization problem is computationally tractable when her utility function is symmetric. In Section~\ref{sec:nonforest}, we prove that the latter is not true if the communication hierarchy is not guaranteed to be a forest. Section~\ref{sec:conclusion}~concludes.

\section{Preliminaries}
\label{sec:prelim}
Similarly to~\cite{arieli2019private}, we consider a Bayesian persuasion setting with one \emph{sender} (she) and $n$ \emph{receivers}. Each receiver $i\in\brackets*{n}$ (he) has a binary \emph{action space} $A_i=\braces*{0,1}$, where $\brackets*{n}:=\braces*{1,\ldots,n}$. For $i\in\brackets*{n}$, denote the action of the $i$-th receiver by $a_i$. Let $S$ denote the set of receivers who choose $a_i=1$. The sender's utility $u\parentheses*{S}$ is specified by some function $u:2^{\brackets*{n}}\to \mathbb{R}_{\ge 0}$. As~\cite{arieli2019private} do, we assume that $u$ is a \emph{monotone} set function, meaning that $S_1\subseteq S_2$ implies $u\parentheses*{S_1}\le u\parentheses*{S_2}$.

Fix a binary \emph{space of states} $\Omega:=\braces*{0,1}$. The actual \emph{state} $\omega$ is drawn from $\Omega$ according to a common-knowledge full-support \emph{prior distribution} $\mu$. As there are only two possible states -- for convenience, we identify $\mu$ with $\Pr_{\mu}[\omega=1]$. That is, $\mu$ is a real number in $\parentheses*{0,1}$. The state $\omega$ is known to the sender, but not to the receivers.

For $i\in\brackets*{n}$, receiver $i$ has utility function $v_i:\Omega\times A_i\to\mathbb{R}$. That is, the utility $v_i\parentheses*{\omega,a_i}$ of receiver $i$ depends both on the state $\omega$ and on his action $a_i$. \emph{In particular, the receivers have no externalities -- i.e., each receiver's utility depends only on the state and that receiver's own action.} We assume that for every $i$, $v_i\parentheses*{\omega,1}-v_i\parentheses*{\omega,0}$ is weakly positive when $\omega=1$ and is strictly negative when $\omega=0$. That is, each receiver $i$ weakly prefers to take the action $a_i=1$ when $\omega=1$ and strictly prefers $a_i=0$ when $\omega=0$.

\paragraph{Forest communication hierarchies.}
Similarly to~\cite{brooks2022information}, we assume that the sender can transmit information to (possibly overlapping) sets of receivers as specified by a \emph{communication hierarchy}. Specifically, there are several \emph{communication channels} between the sender and some sets of receivers. Given two receivers $i_1\neq i_2$, we say that $i_1$ (strictly) \emph{dominates} $i_2$ if $i_1$ observes all the communication channels that $i_2$ does, but there is at least one communication channel observed by $i_1$, but not by $i_2$. Note that the domination binary relation is a partial order over the set of receivers. The \emph{communication hierarchy} is a directed graph $G=\parentheses*{\brackets*{n},E}$, where the vertices are the receivers, and each $\parentheses*{i_1,i_2}\in E$ corresponds to a receiver $i_1$ dominating $i_2$ s.t. there is no receiver $i_3\neq i_1,i_2$ between $i_1$ and $i_2$ in the domination partial order. Throughout the paper, except for Section~\ref{sec:nonforest}, we assume that $G$ is a rooted directed forest. The \emph{depth} of a receiver $i$ is the number of edges in the unique directed path from the root of the tree component of $G$ containing $i$ to $i$ himself. The \emph{depth of $G$}, denoted $h=h\parentheses*{G}$, is the maximum depth of any receiver.

\paragraph{The signaling scheme.}
Let $M$ be some abstract set of $\emph{signals}$. As is standard in Bayesian persuasion, we assume that at the beginning of the sender-receivers interaction, the sender must commit to a \emph{signaling scheme} $\Sigma:\Omega\to\Delta\parentheses*{M^n}$, where $\Delta\parentheses*{\cdot}$ is the set of all probability distributions over some set. The signaling scheme specifies for each possible state $\omega$ the distribution of the signals sent to all the receivers when the state is $\omega$. The difference from the standard model is that each receiver $i$ observes not only the message sent directly to him, but also the messages sent to all the receivers dominated by $i$ -- i.e., all the descendants of $i$ in the communication hierarchy $G$. Upon observing the signals, each receiver $i$ performs a Bayesian update to his belief about the distribution of the state, resulting in a \emph{posterior} probability $p_i=p_i\parentheses*{\sigma_i}$ for $\omega=1$, where $\sigma_i$ is a concatenation of the signals that receiver $i$ observes. In a formula,
$$
p_i=p_i\parentheses*{\sigma_i}=\frac{\mu\Pr_{\Sigma\parentheses*{\omega=1}}[\sigma_i]}{\mu\Pr_{\Sigma\parentheses*{\omega=1}}[\sigma_i]+\parentheses*{1-\mu}\Pr_{\Sigma\parentheses*{\omega=0}}[\sigma_i]}.
$$
Receiver $i\in\brackets*{n}$ takes action $1$ iff $\mathbb{E}_{\omega\sim p_i}\brackets*{v_i\parentheses*{\omega,1}}\ge \mathbb{E}_{\omega\sim p_i}\brackets*{v_i\parentheses*{\omega,0}}$. That is, each receiver takes his best response action, with ties, as is standard, broken in the sender's favour.

\paragraph{The information structure.}
Let $p=\parentheses*{p_1,\ldots,p_n}$ be the \emph{joint posterior}. Note that only the joint posterior $p$ affects the interaction outcome, not the concatenations of signals $\sigma_1,\ldots,\sigma_n$ themselves. Moreover, a signaling scheme $\Sigma$ naturally induces a distribution $\pi=\pi\parentheses*{\Sigma}\in\Delta\parentheses*{\parentheses*{\Delta\parentheses*{\Omega}}^n}$ over joint posteriors (unconditional on the state).\footnote{Here we identify the joint posterior $p=\parentheses*{\Pr_{p_1}[\omega=1],\ldots,\Pr_{p_n}[\omega=1]}$ with the corresponding distribution over $\Omega^n=\braces*{0,1}^n$.} We call such $\pi$ the \emph{information structure} induced by the signaling scheme $\Sigma$.

For $i\in\brackets*{n}$, let $\pi_i$ be the projection of $\pi$ to $\Delta\parentheses*{p_i}$. It is well-known that a necessary condition for an information structure $\pi$ to be inducible by \emph{some} signaling scheme is Bayes-plausibility -- namely, we must have $\mathbb{E}_{p_i\sim\pi}\brackets*{p_i}=\mu$ for every $i\in\brackets*{n}$; moreover, for $n=1$ this condition is also sufficient~\cite{blackwell1950comparison,Blackwell53,aumann1995repeated,kamenica2011bayesian}. For a general number of receivers and communication hierarchy, it might be complicated to determine whether a given information structure $\pi$ can be induced by a signaling scheme. However,~\cite{brooks2022information} prove the following complete characterization for \emph{forest} communication~hierarchies.

\begin{fact}[\citet{brooks2022information}]
\label{fact:inducible}
    Let $G$ be a forest communication hierarchy and $\pi_1,\ldots,\pi_n\in\Delta\parentheses*{\Delta\parentheses*{\Omega}}$. Then there exists a distribution over joint posteriors $\pi\in\Delta\parentheses*{\parentheses*{\Delta\parentheses*{\Omega}}^n}$ that is inducible by a signaling scheme and has the respective marginals $\pi_1,\ldots,\pi_n$ iff the following two conditions hold:
    \begin{itemize}
    \item For every $i\in\brackets*{n}$, $\mathbb{E}_{p_i\sim\pi_i}\brackets*{p_i}=\mu$ (Bayes-plausibility).
    \item For every $\parentheses*{i_1,i_2}\in E$, $\pi_{i_1}$ is a \emph{mean-preserving spread} of $\pi_{i_2}$.\footnote{Given $\nu_1,\nu_2\in\Delta\parentheses*{\Delta\parentheses*{\Omega}}$ and corresponding random variables $X_{\nu_1},X_{\nu_2}$, we say that $\nu_1$ is a \emph{mean-preserving spread} of $\nu_2$ if $X_{\nu_1},X_{\nu_2}$ have the same expected value, and there exists a random variable $Y$ s.t.~$X_{\nu_1}\equiv X_{\nu_2}+Y$ and the expectation of $Y$ conditional on any value of $X_{\nu_2}$ is $0$ (see, e.g.,~\cite{Mean}).}
    \end{itemize}
\end{fact}

Given this characterization, from now on we shall focus on the information structure $\pi$ and its marginals $\pi_1,\ldots,\pi_n$ rather than the signaling scheme $\Sigma$. Note further that $\Sigma$ induces distributions $\pi^0, \pi^1\in\Delta\parentheses*{\parentheses*{\Delta\parentheses*{\Omega}}^n}$ conditional on $\omega=0$ and $\omega=1$, respectively.

\paragraph{The full sender-receivers interaction summarized.}
\begin{enumerate}
    \item The sender commits to a signaling scheme $\Sigma$ that induces some information structure $\pi$.
    \item The state $\omega$ is drawn from the prior distribution $\mu$ and is reported to the sender.
    \item The joint posterior $p=\parentheses*{p_1,\ldots,p_n}$ is drawn from $\pi^{\omega}$.
    \item Each receiver $i\in\brackets*{n}$ computes his posterior $p_i$ and takes a best-response action $a_i$, with ties broken in the sender's favour.
    \item The sender gets utility of $u\parentheses*{S}$, where $S=\braces*{i\in\brackets*{n}:a_i=1}$, and each receiver $i\in\brackets*{n}$ gets utility of $v_i\parentheses*{\omega,a_i}$.
\end{enumerate}

\paragraph{Representation of the utility functions.}
We assume that the sender has oracle access to her utility function $u\parentheses*{S}$, while each receiver's utility function $v_i\parentheses*{\omega,a_i}$ is given by an explicit formula.

\paragraph{The sender's optimization problem.}
Similarly to most Bayesian persuasion papers, we focus on the sender's design of an information structure maximizing her expected utility. That is, our goal is to compute an information structure $\pi$ maximizing $V\parentheses*{\pi}:=\mathbb{E}_{p\sim\pi}\brackets*{u\parentheses*{S\parentheses*{p}}}$, where $S\parentheses*{p}$ is the set of all the receivers choosing action $1$ when the joint posterior is $p$. We call $V\parentheses*{\pi}$ the sender's \emph{value} under the information structure $\pi$, and we call $V^{*}:=\max_{\pi} V\parentheses*{\pi}$ the sender's \emph{optimal value}. Note that the maximum is achieved since $V\parentheses*{\pi}$ is an upper semicontinuous functional of $\pi$ and the space of feasible arguments $\pi$ is compact. An information structure $\pi$ achieving $V^{*}$ is called~\emph{optimal}.

\paragraph{Facilitating the problem.}
It follows directly from the setting definition that each receiver $i\in\brackets*{n}$ has a \emph{persuasion threshold} $t_i\in (0,1]$ s.t.~$a_i=1$ iff $p_i\ge t_i$. Specifically, $t_i$ is given by the condition: $t_i v_i\parentheses*{\omega=1, a_i=1}+\parentheses*{1-t_i}v_i\parentheses*{\omega=0, a_i=1}=t_i v_i\parentheses*{\omega=1, a_i=0}+\parentheses*{1-t_i}v_i\parentheses*{\omega=0, a_i=0}$. For simplicity of presentation, set $t_0:=0$ and $t_{n+1}:=1$. Assume w.l.o.g.~that $t_1\le t_2\le\ldots\le t_n$. Let $\Theta:=\braces*{t_0,\ldots,t_{n+1}}$.

Recall that for a distribution $\nu$ taking values in a countable set $C$, the \emph{support} of $\nu$ is $\supp\parentheses*{\nu}:=\braces*{x\in C:\nu\parentheses*{x}>0}$.\footnote{For a general compact set $C$, $\supp\parentheses*{\nu}$ is the minimal closed set s.t.~$\Pr_{x\sim\nu}[x\in \supp\parentheses*{\nu}]=1$.} The following lemma allows us to discretize the sender's optimization problem to information structures supported on the grid $\Theta^n$. This property will be helpful in analyzing the sender's optimization problem throughout the paper.

\begin{lemma}
    \label{lem:grid}
    Fix a forest communication hierarchy $G$. Then for any information structure $\pi$, there exists an information structure $\pi'$ with $\supp\parentheses*{\pi_i'}\subseteq  \Theta$ for any $i\in\brackets*{n}$ s.t.~$\mathbb{E}_{p\sim\pi'}\brackets*{u\parentheses*{S\parentheses*{p}}}\ge \mathbb{E}_{p\sim\pi}\brackets*{u\parentheses*{S\parentheses*{p}}}$.
\end{lemma}

The proof is based on iteratively (over the receivers) spreading the probability mass assigned by $\pi_i$ to each $q_i\in\supp\parentheses*{\pi_i}\setminus\Theta$ in a mean-preserving way across the closest elements of $\Theta$ on either side of $q_i$; such a spread requires applying a correction procedure making sure that the restriction of $\pi$ to receivers dominating $i$ is still inducible without suffering any loss in expected sender's utility; the correction argument is close in spirit to the repeated refinement of signals argument of~\cite{brooks2022information} in Fact~\ref{fact:inducible} proof. For the full proof, see Appendix~\ref{app:prelim}.

\begin{notation}
\label{not}
Let $0=\tau_0<\tau_1<\ldots<\tau_m=1$ be s.t.~$\Theta=\braces*{\tau_0,\ldots,\tau_m}=\braces*{t_0,\ldots,t_{n+1}}$.
\begin{itemize}
    \item  For $i=1,\ldots,n$, let $D_i$ be the set of receivers dominated by receiver $i$ together with receiver $i$ himself. Denote further $d_i:=\absolute*{D_i}$.
    \item For $i=1,\ldots,n$, let $\Delta_i:=\braces*{\tau_0,\ldots,\tau_m}^{d_i}$ be the set of all possible vectors of grid-valued marginal posteriors of receivers in $D_i$.
    \item For a joint posterior $p$, denote by $p|_{D_i}\in \Delta_i$ its projection on $D_i$.
    \item For $q\in \Delta_i$ and $j\in D_i$, define $q_j$ to be the corresponding marginal posterior of receiver $j$.
    \item Let $L:=\braces*{\bar{l}=\parentheses*{l_1,\ldots,l_n}:\; 0\le l_1,\ldots,l_n\le m}$. For $\bar{l}\in L$, define $S\parentheses*{\bar{l}}:=\braces*{i\in [n]:\; \tau_{l_i}\ge t_i}$ and $p\parentheses*{\bar{l}}:=\parentheses*{\tau_{l_1},\ldots,\tau_{l_n}}$.
\end{itemize}
\end{notation}

\section{Warm-Up: Tractability for Modular Sender's Utility}
\label{sec:mod}
In this section, we show that computing an optimal information structure under a forest communication hierarchy is computationally tractable when the sender's utility function $u$ is \emph{modular} -- i.e.,  $u\parentheses*{S}=\sum_{i\in S} \lambda_i$ for some rational weights $\lambda_1,\ldots,\lambda_n\ge 0$. This is a partial generalization of the computational tractability for supermodular sender's utility when the communication is private that was shown by~\cite{arieli2019private}.

\begin{proposition}
    \label{pro:linear}
    For a modular sender's utility function $u\parentheses*{S}=\sum_{i\in S} \lambda_i$ for some rational weights $\lambda_1,\ldots,\lambda_n\ge 0$, there is a weakly polynomial-time algorithm for computing an optimal signaling scheme for every forest communication hierarchy $G$.
\end{proposition}

The proof utilizes Fact~\ref{fact:inducible} and Lemma~\ref{lem:grid}. The key property of a modular sender's utility is that the marginal information structures $\pi_1,\ldots,\pi_n$ suffice to determine the sender's value $V\parentheses*{\pi}$ regardless of the way they are coupled into a joint information structure $\pi$. Since Bayes-plausibility and mean-preserving spread constraints are linear, and Lemma~\ref{lem:grid} allows us to restrict attention to a finite grid, the proof uses a natural linear program. For details, see Appendix~\ref{app:modular}.

\section{The Supermodular and the Submodular Cases}
\label{sec:supersub}
We start with the following key lemma that does not impose any assumptions on the sender's utility $u:2^{\brackets*{n}}\to \mathbb{R}_{\ge 0}$ beyond monotonicity.

\begin{lemma}
    \label{lem:lp}
Fix a forest communication hierarchy $G$. Then the sender's optimal value equals the value of the optimal solution of the following LP:

\begin{equation}
\label{lp:1}
\begin{aligned}
    \text{max}\;\;\;\;&\sum_{\bar{l}\in L} u\parentheses*{S\parentheses*{\bar{l}}} \cdot \parentheses*{\mu x_{\bar{l}}^1+\parentheses*{1-\mu} x_{\bar{l}}^0}\\
    \text{s.t.}\;\;\;\;\;\;&\sum_{\bar{l}\in L:\; p\parentheses*{\bar{l}}|{D_i}=q}\parentheses*{\parentheses*{1-q_i}\mu x_{\bar{l}}^{1}- q_i\parentheses*{1-\mu} x_{\bar{l}}^{0}}=0\;\;\;\;\;\;\;\;\; i=1,\ldots,n,\;\; q\in \Delta_i\\
    &\sum_{\bar{l}\in L} x_{\bar{l}}^{w} = 1 \;\;\;\;\;\;\;\;\;\;\;\;\;\;\;\;\;\;\;\;\;\;\;\;\;\;\;\;\;\;\;\;\;\;\;\;\;\;\;\;\;\;\;\;\;\;\;\;\;\;\;\;\;\;\;\;\;\;\;\;\;\;\;\; w=0,1\\
    &x_{\bar{l}}^w\ge 0\;\;\;\;\;\;\;\;\;\;\;\;\;\;\;\;\;\;\;\;\;\;\;\;\;\;\;\;\;\;\;\;\;\;\;\;\;\;\;\;\;\;\;\;\;\;\;\;\;\;\;\;\;\;\;\;\;\;\;\;\;\;\;\;\;\;\;\;\;\; \bar{l}\in L,\;\; w=0,1.
    \end{aligned}
\end{equation}

\end{lemma}

This lemma is a generalization of the characterization of all admissible joint information structures given the distributions of marginal posteriors that was derived by~\cite{arieli2022persuasion} for the case of private persuasion. However, the proof presents additional technical challenges. Indeed, when a certain signal is observed by more than one receiver it is no longer true that a set of admissible marginal posteriors conditional on a certain fixed value of the state $\omega$ can be coupled in an arbitrary way. To capture the dependencies between the information observed by dominating pairs of receivers, we have to introduce a separate constraint for every subtree of the given forest $G$ and any possible values from the grid $\Theta$ assigned to the marginal posteriors of the receivers in this subtree.

Both the number of constraints and the number of variables in LP~(\ref{lp:1}) are non-polynomial in $n$. Thus, in order to derive positive computational results, one has to build on some features of a specific class of utility functions $u$. Still, Lemma~\ref{lem:lp} can be of independent interest since it provides necessary and sufficient conditions for inducibility of an information structure supported on the grid $\Theta^n$, where $x_{\bar{l}}^w$ represents the probability to induce the joint posterior $p\parentheses*{\bar{l}}$ conditional on $\omega=w$. For the proof of Lemma~\ref{lem:lp}, see Appendix~\ref{app:lp}.

\subsection{Hardness of the Supermodular Case}
\label{sub:super}
In this subsection, we show that unlike in the private persuasion case, the problem of maximizing the sender's value under a forest hierarchy $G$ is $NP$-hard when the sender's utility is supermodular. Moreover, there is no FPTAS unless $P=NP$, and it is true even when $G$ is a disjoint union of chain components each of which has just $3$ receivers.

\begin{theorem}
    \label{thm:super_hard}
    Unless $P=NP$, there is no FPTAS for approximating the sender's optimal value with a monotone supermodular sender's utility and a forest communication hierarchy $G$ even if $G$ is a disjoint union of chains of $3$ receivers.
\end{theorem}

Intuitively, while under private persuasion in the supermodular case the sender aims to positively correlate the distributions over marginal posteriors of different receivers as much as possible, the presence of chains significantly complicates the sender's decision problem. To prove hardness, we reduce from the max-cut problem. Specifically, given an  undirected graph $\parentheses*{V,E'}$ with $\absolute*{V}=k$ and $\absolute*{E'}=e$, we construct a Bayesian persuasion instance by picking a uniform prior and creating a chain $v_1 \implies v_2\implies v_3$ for every $v\in V$. Every $v_1$ will have the same threshold $t_{v_1}=\alpha$ which will be close to $1$ -- i.e., the most informed receivers will be the hardest to persuade; we shall further choose $t_{v_2}=1/4$ for every $v\in V$, making the middle node in each chain very easy to persuade, and $t_{v_3}=3/4$ for every $v\in V$, making the least informed receivers easier to persuade than the most informed ones $v_1$ but harder compared to $v_2$.

The sender's utility $u$ will be defined as $u\parentheses*{S}:=M_1 u_1\parentheses*{S} + M_2 u_2\parentheses*{S} + u_3\parentheses*{S}$ for some rational $M_1,M_2>0$ which will be chosen s.t.:

\begin{itemize}
    \item Both $M_1$ and $M_2$ have a numerator and a denominator which are $\poly\parentheses*{k}$.
    \item Maximizing $u$ over information structures supported on $\Theta^n$ (which is w.l.o.g.~by Lemma~\ref{lem:grid}) is equivalent to a lexicographic maximization -- i.e., every information structure $\pi$ with $\supp\parentheses*{\pi}\subseteq\Theta^n$ maximizing $\mathbb{E}\brackets*{u}$ maximizes $\mathbb{E}\brackets*{u_1\parentheses*{S}}$; maximizes $\mathbb{E}\brackets*{u_2\parentheses*{S}}$ subject to maximizing $\mathbb{E}\brackets*{u_1\parentheses*{S}}$; and maximizes $\mathbb{E}\brackets*{u_3\parentheses*{S}}$ subject to the two aforementioned maximizations.
\end{itemize}

We shall choose $u_1$ to be a modular function s.t.~maximizing it will ensure that the restriction of $\pi$ on each chain component of $G$ will fix the distribution over $\parentheses*{p_{v_1},p_{v_3}}$, though the correlation between the restrictions of $\pi$ to different components still remains unrestricted at this point, and the distributions over $p_{v_2}$ have a lot of freedom. Next, we shall pick a supermodular $u_2$ that, subject to maximizing $\mathbb{E}\brackets*{u_1\parentheses*{S}}$, ensures that maximizing $\mathbb{E}\brackets*{u_2\parentheses*{S}}$ forces the distributions over $\parentheses*{p_{v_1},p_{v_3}}$ for different $v\in V$ to be correlated s.t.~$X_v:=\mathbb{I}_{v_1\in S}+\mathbb{I}_{v_3\in S}$ is the same for every $v\in V$ w.p.~$1$.

Finally, we shall pick $u_3\parentheses*{S}:=\sum_{\{v,w\}\in E'} \parentheses*{\mathbb{I}_{v_1,w_3\in S}+\mathbb{I}_{v_3,w_1\in S}}$, which is supermodular. Our construction will ensure that the lexicographic maximization will be reduced, for an appropriate choice of $\alpha>0$, to maximizing the number of edges $\braces*{v,w}\in E'$ s.t.~$\parentheses*{p_{v_1},p_{v_2},p_{v_3}}=\parentheses*{\alpha ,1/4,1/4}$ and $\omega=0$ happens iff $\parentheses*{p_{w_1},p_{w_2},p_{w_3}}=\parentheses*{0,3/4,3/4}$; moreover, $\parentheses*{p_{w_1},p_{w_2},p_{w_3}}=\parentheses*{\alpha ,1/4,1/4}$ and $\omega=0$ happens iff $\parentheses*{p_{v_1},p_{v_2},p_{v_3}}=\parentheses*{0,3/4,3/4}$. Since this is equivalent to finding a max-cut, the reduction follows.

Ensuring that the maximization of $\mathbb{E}\brackets*{u\parentheses*{S}}$ forces a lexicographic maximization of $\mathbb{E}\brackets*{u_1\parentheses*{S}}$, $\mathbb{E}\brackets*{u_2\parentheses*{S}}$, $\mathbb{E}\brackets*{u_3\parentheses*{S}}$ imposes significant technical challenges.\footnote{While the high-level idea of our lexicographic optimization was partially inspired by~\cite{behringer1977lexicographic,sherali1982equivalent,isermann1982linear,dughmi2016algorithmic}, and the choice of $M_1$ resembles general approaches in linear programming (see, e.g.,~\cite{schrijver1998theory}), our construction does not follow from these sources and combines multiple intricate ideas.} To force maximization of $\mathbb{E}\brackets*{u_1\parentheses*{S}}$, we exploit the modularity of $u_1$, which means that maximizing $\mathbb{E}\brackets*{u_1\parentheses*{S}}$ is equivalent to maximizing over each component of $G$ separately.

If $\mathbb{E}\brackets*{u_1\parentheses*{S}}$ is not maximized by some information structure $\pi$, then there exists $v\in V$ s.t.~$E_v$, the expectation of the summand corresponding to the $v$-th chain component in $\mathbb{E}\brackets*{u_1\parentheses*{S}}$, is not maximized. Note that maximizing $E_v$ is optimizing a linear objective over an $O(1)$-sized convex polyhedron $P$ that is provided by Lemma~\ref{lem:lp}. Thus, there exists a global constant $C_v=O(1)$ equal to the minimal difference between two distinct values of $E_v$ at vertices of $P$. It can be shown by a straightforward computation that improving $E_v$ to its maximum possible value, while preserving the validity of $\pi$, increases $\mathbb{E}\brackets*{u_1\parentheses*{S}}$ by at least $C_v \beta$, where $\beta$ depends on the representation of the point corresponding to the restriction of $\pi$ to the $v$-th chain component as a convex combination of the vertices of $P$. We further show that the expectations of $u_2,u_3$ are bounded by an expression that depends linearly on $\beta$, and thus an appropriate choice of $M_1$ implies that maximizing $\mathbb{E}\brackets*{u\parentheses*{S}}$ forces maximization of $\mathbb{E}\brackets*{u_1\parentheses*{S}}$.

The choice of $M_2$ is more ad hoc compared to $M_1$ and is based on the specific forms that $\mathbb{E}\brackets*{u_2\parentheses*{S}}$ and $\mathbb{E}\brackets*{u_3\parentheses*{S}}$ take when $\mathbb{E}\brackets*{u_1\parentheses*{S}}$ is maximized. Namely, we bound the possible gain in $\mathbb{E}\brackets*{u_3\parentheses*{S}}$ compared to the loss in $\mathbb{E}\brackets*{u_2\parentheses*{S}}$ when tampering with the correlations between the restrictions of $\pi$ to different chain components.

\begin{proof}[Proof of Theorem~\ref{thm:super_hard}]
We reduce from the decision version of the max-cut problem, which is known to be NP-complete~\cite{karp1972reducibility}.

\paragraph{Step 1: Constructing the Bayesian persuasion instance.} Suppose that we are given a simple undirected graph $\parentheses*{V,E'}$ with $\absolute*{V}=k>1$, $\absolute*{E'}=e>1$. Let $\mc\parentheses*{V,E'}$ be the size of the maximum cut in the given graph. Construct a persuasion instance as follows: 
    \begin{itemize}
        \item For every vertex $v\in V$, there will be $3$ receivers $v_1,v_2,v_3$ arranged in a chain -- i.e., $v_2$ dominates $v_3$ and $v_1$ dominates $v_2,v_3$; the receivers corresponding to distinct vertices are incomparable. In particular, $n=3k$.
        \item For every $v\in V$, set $t_{v_1}=\alpha$ for $\alpha>3/4$ that we shall choose later, $t_{v_2}=1/4$, and $t_{v_3}=3/4$.
        \item Let $\mu$ be uniform.
        \item The sender's utility $u$ will be constructed gradually through multiple steps.
    \end{itemize}

\paragraph{Step 2: Optimizing a simple modular objective.}
Define $u_1\parentheses*{S}:=\sum_{v\in V}\parentheses*{\mathbb{I}_{v_1\in S}+2\mathbb{I}_{v_2\in S}+2\mathbb{I}_{v_3\in S}}$.

Fix some $v\in V$. From Markov's inequality and Bayes-plausibility of the marginal posteriors, we get that under any information structure (where the probability is over the information structure):

\begin{equation}
    \label{eq:Markov}
    \Pr[v_1\in S]\le\frac{1}{2\alpha},
\end{equation}

with equality iff $\supp\parentheses*{p_{v_1}}=\braces*{0,t_{v_1}}$. Next, we claim that the following inequality holds under any information structure and for any $q\in\supp\parentheses*{p_{v_3}}$:

\begin{equation}
    \label{eq:NoRoot}
    2\mathbb{I}_{v_3\in S}+2\Pr[v_2\in S|p_{v_3}=q]\le 1+4q.
\end{equation}

Indeed, since $\mathbb{E}\brackets*{p_{v_2}|p_{v_3}}=p_{v_3}$, Markov's inequality implies $\Pr[v_2\in S|p_{v_3}=q]=\Pr[p_{v_2}\ge 1/4|p_{v_3}=q]\le 4q$. Since probability is always at most $1$, it suffices to prove:

\begin{equation*}
    2\mathbb{I}_{v_3\in S}+2\min\braces*{1,4q}\le 1+4q.
\end{equation*}

Let us consider the following three cases:

\begin{enumerate}
    \item If $q<t_{v_2}=1/4<t_{v_3}=3/4$, then $LHS=8q<1+4q$.
    \item If $t_{v_2}=1/4\le q<t_{v_3}=3/4$, then $LHS=2\le 1+4q$, with equality iff $q=1/4$.
    \item Finally, if $t_{v_2}=1/4<t_{v_3}=3/4\le q$, then $LHS=4\le 1+4q$, with equality iff $q=3/4$.
\end{enumerate}

Thus,~(\ref{eq:NoRoot}) indeed holds, with equality iff $\supp\parentheses*{p_{v_3}}=\braces*{t_{v_2},t_{v_3}}$ and $p_{v_2}\ge t_{v_2}=1/4$ w.p.~$1$. Taking the expectation over the marginal distributions of $p_{v_2},p_{v_3}$ in~(\ref{eq:NoRoot}) and adding~(\ref{eq:Markov}) implies that for every $v\in V$ and every information structure:

\begin{equation*}
    \mathbb{E}\brackets*{\mathbb{I}_{v_1\in S}+2\mathbb{I}_{v_2\in S}+2\mathbb{I}_{v_3\in S}}\le \frac{1}{2\alpha} + 1 + 4\cdot \frac{1}{2} = \frac{6\alpha+1}{2\alpha}.
\end{equation*}

Summing over $v\in V$ yields:

\begin{equation}
\label{eq:component1}
    \mathbb{E}\brackets*{u_1\parentheses*{S}}\le \frac{6\alpha+1}{2\alpha}k,
\end{equation}

with equality iff for every $v\in V$, $\supp\parentheses*{p_{v_1}}=\braces*{0,t_{v_1}=\alpha}$ and $\supp\parentheses*{p_{v_3}}=\braces*{t_{v_2}=1/4,t_{v_3}=3/4}$ and $p_{v_2}\ge t_{v_2}=1/4$ w.p.~$1$. Note that by Bayes-plausibility, the conditions on the support mean that $p_{v_1}$ gets value $\alpha$ w.p.~$1/2\alpha$ and value $0$ w.p.~$1-1/2\alpha$, while $p_{v_3}$ is uniform over $1/4,3/4$. Note that if we spread the mass of $1/2$ from each of $1/4,3/4$ in a mean-preserving way, the most we can assign to $\alpha$ is achieved by distributing the mass between $\alpha$ and $0$, which will result in $\alpha$ being assigned exactly the mass of $1/8\alpha+3/8\alpha=1/2\alpha$. Therefore, an information structure $\pi$ supported on $\Theta^n$ achieves the upper bound of $\frac{6\alpha+1}{2\alpha}k$ iff it satisfies the following for every $v\in V$:

\begin{enumerate}
\item w.p. $3/8\alpha$, $\parentheses*{p_{v_1},p_{v_3}}=\parentheses*{\alpha,3/4}$;
\item w.p. $1/8\alpha$, $\parentheses*{p_{v_1},p_{v_3}}=\parentheses*{\alpha,1/4}$;
\item w.p. $1/2-3/8\alpha$, $\parentheses*{p_{v_1},p_{v_3}}=\parentheses*{0,3/4}$;
\item w.p. $1/2-1/8\alpha$, $\parentheses*{p_{v_1},p_{v_3}}=\parentheses*{0,1/4}$.
\item w.p. $1$, $p_{v_2}\ge 1/4$.
\end{enumerate}

\paragraph{Step 3: Coupling the marginals.} Suppose that there is a way to enforce that the information structure $\pi$ is supported on $\Theta^n$ and maximizes $\mathbb{E}\brackets*{u_1\parentheses*{S}}$. Define the following function:

\begin{equation}
    u_2\parentheses*{S}:=\sum_{\{v,w\}\subseteq V:\; v\neq w} \parentheses*{\mathbb{I}_{v_1\in S}+\mathbb{I}_{v_3\in S}}\parentheses*{\mathbb{I}_{w_1\in S}+\mathbb{I}_{w_3\in S}}.
\end{equation}

Note that the summand corresponding to $v,w$ in the definition of $u_2$ equals $\mathbb{I}_{v_1,w_1\in S}+\mathbb{I}_{v_1,w_3\in S}+\mathbb{I}_{v_3,w_1\in S}+\mathbb{I}_{v_3,w_3\in S}$, and thus $u_2$ is supermodular (as a sum of supermodular functions).

Let $X_v:=\mathbb{I}_{v_1\in S}+\mathbb{I}_{v_3\in S}$ ($v\in V$). Then we have for $v\ne w$:

\begin{equation*}
    \mathbb{E}\brackets*{X_v X_w}=\frac{\mathbb{E}\brackets*{X_v^2}+\mathbb{E}\brackets*{X_w^2}-\mathbb{E}\brackets*{\parentheses*{X_v - X_w}^2}}{2}.
\end{equation*}

Thus, among the $\pi$s that maximize $\mathbb{E}\brackets*{u_1\parentheses*{S}}$, those that maximize $\mathbb{E}\brackets*{u_2\parentheses*{S}}$ are precisely the $\pi$s for which $X_v$ is equal for all $v\in V$ w.p.~$1$. Moreover, in this case, we have for every $v\ne w$: $\mathbb{E}\brackets*{X_v X_w}=\mathbb{E}\brackets*{X_v^2}=\mathbb{E}\brackets*{X_w^2}$. By Points $1-4$ from the previous step and conditional probability, we get that for $v\in V$:

\begin{equation*}
    \mathbb{E}\brackets*{X_v^2}=\mathbb{E}\brackets*{\mathbb{I}_{v_1\in S}+\mathbb{I}_{v_3\in S}+2\mathbb{I}_{v_1,v_3\in S}}=\frac{1}{2\alpha}+\frac{1}{2}+2\cdot\frac{3}{8\alpha}=\frac{2\alpha+5}{4\alpha}.
\end{equation*}
Thus, for every $\pi$ satisfying Points $1-4$ from the previous step, we have:

\begin{equation}
    \label{eq:coupling}
    \mathbb{E}\brackets*{u_2\parentheses*{S}}\le \frac{\parentheses*{2\alpha+5}k(k-1)}{8\alpha},
\end{equation}

with equality iff all the $X_v$s are equal w.p.~$1$.

\paragraph{Step 4: Translating max-cut into a supermodular utility function.}

Assume now that one can enforce that $\pi$ is supported on $\Theta^n$ and maximizes $\mathbb{E}\brackets*{u_2\parentheses*{S}}$ among those $\pi$ that maximize $\mathbb{E}\brackets*{u_1\parentheses*{S}}$. Call such $\pi$ \emph{good}. Define the following supermodular function:

\begin{equation}\label{eq:u3}
    u_3\parentheses*{S}:=\sum_{\{v,w\}\in E'} \parentheses*{\mathbb{I}_{v_1,w_3\in S}+\mathbb{I}_{v_3,w_1\in S}}.
\end{equation}

Note that in Points $1,2,3,4$ from Step 2, we have $X_v=2,1,1,0$, respectively. Thus, for every good $\pi$, we have for any $v\ne w$ that $\mathbb{I}_{v_1,w_3\in S}+\mathbb{I}_{v_3,w_1\in S}=2$ if Point 1 from Step 2 holds for every $v'\in V$. Moreover, $\mathbb{I}_{v_1,w_3\in S}+\mathbb{I}_{v_3,w_1\in S}=0$ if Point 4 holds for every $v'\in V$. Otherwise, $\mathbb{I}_{v_1,w_3\in S}+\mathbb{I}_{v_3,w_1\in S}\in \{0,1\}$, with the value being $1$ iff Point 2 holds for exactly one of $v,w$, while Point 3 holds for the other one.

Note that Points $3$ and $4$ can only hold when the state is $\omega=0$, since the posterior of one of the receivers is $0$. The probability that $\omega=0$ and Point 2 holds is the probability that $\omega=0$, the least informed receiver in the chain has the posterior of $1/4$, but Point $4$ does not hold. Therefore, it equals $\frac{3}{8}-\parentheses*{\frac{1}{2}-\frac{1}{8\alpha}}=\frac{1}{8\alpha}-\frac{1}{8}$. To make sure that the probability that $\omega=0$ and Point 2 holds equals the probability that Point 3 holds, we have to choose $\alpha$ s.t.~$\frac{1}{8\alpha}-\frac{1}{8}=\frac{1}{2}-\frac{3}{8\alpha}$, i.e.~$\alpha=\frac{4}{5}$.

Note that for any realization of signals s.t.~for each vertex, either Point 2 or Point 3 holds, we have the following property: The edges connecting vertices corresponding to different Points are a cut of $\parentheses*{V,E'}$. Thus, the theoretical upper bound on $\mathbb{E}\brackets*{u_3\parentheses*{S}}$ for good $\pi$ is:

\begin{equation}
    \label{eq:max_cut}
    \frac{3}{8\alpha}\cdot 2\cdot e + 2\parentheses*{\frac{1}{2}-\frac{3}{8\alpha}}\cdot \mc\parentheses*{V,E'}=\frac{15}{16}e+\frac{\mc\parentheses*{V,E'}}{16}.
\end{equation}

Moreover, given a maximum cut $\parentheses*{T_1,T_2}$, this bound is achievable by the following good $\pi$:

\begin{enumerate}
\item w.p. $3/8\alpha=15/32$, $\parentheses*{p_{v_1},p_{v_2},p_{v_3}}=\parentheses*{\alpha=4/5,3/4,3/4}$ for every $v\in V$ (the conditional probability on $\omega$ to induce this joint posterior is $3/16$ when $\omega=0$ and $3/4$ when $\omega=1$);
\item w.p. $1/8\alpha-1/32=1/8$, $\parentheses*{p_{v_1},p_{v_2},p_{v_3}}=\parentheses*{\alpha=4/5,1/4,1/4}$ for every $v\in V$ (conditional on $\omega$: $0$ when $\omega=0$ and $1/4$ when $\omega=1$);
\item w.p. $1/2-3/8\alpha=1/32$, $\parentheses*{p_{v_1},p_{v_2},p_{v_3}}=\parentheses*{0,3/4,3/4}$ for $v\in T_1$ and $\parentheses*{p_{v_1},p_{v_2},p_{v_3}}=\parentheses*{4/5,1/4,1/4}$ for $v\in T_2$ (conditional on $\omega$: $1/16$ when $\omega=0$ and $0$ when $\omega=1$);
\item w.p. $1/2-3/8\alpha=1/32$, $\parentheses*{p_{v_1},p_{v_2},p_{v_3}}=\parentheses*{4/5,1/4,1/4}$ for $v\in T_1$ and $\parentheses*{p_{v_1},p_{v_2},p_{v_3}}=\parentheses*{0,3/4,3/4}$ for $v\in T_2$ (conditional on $\omega$: $1/16$ when $\omega=0$ and $0$ when $\omega=1$);
\item w.p. $1/2-1/8\alpha=11/32$, $\parentheses*{p_{v_1},p_{v_2},p_{v_3}}=\parentheses*{0,1/4,1/4}$ for every $v\in V$ (conditional on $\omega$: $11/16$ when $\omega=0$ and $0$ when $\omega=1$).
\end{enumerate}

\paragraph{Step 5: Defining the sender's utility and completing the reduction.} Set $u\parentheses*{S}:=M_1 u_1\parentheses*{S} + M_2 u_2\parentheses*{S} + u_3\parentheses*{S}$ for some rational $M_1,M_2>0$. Let $V^{*}$ be the optimum value of the Bayesian persuasion problem. We shall show that there exist $M_1,M_2>k^2$ s.t.:

\begin{enumerate}
    \item $M_1,M_2$ have a numerator and a denominator $\poly\parentheses*{k}$ in their reduced form.
    \item Every information structure $\pi$ supported on $\Theta^n$ which is optimal must exactly maximize $\mathbb{E}\brackets*{u_1\parentheses*{S}}$ and maximize $\mathbb{E}\brackets*{u_2\parentheses*{S}}$ subject to maximizing $\mathbb{E}\brackets*{u_1\parentheses*{S}}$.
\end{enumerate}

We first note that if such $M_1,M_2$ exist, we are done. Indeed, suppose that there exists an FPTAS for the Bayesian persuasion problem. We have: $0\le \mathbb{E}\brackets*{u_1\parentheses*{S}}< 15k$, $0\le \mathbb{E}\brackets*{u_2\parentheses*{S}}<4k^2$, $0\le \mathbb{E}\brackets*{u_3\parentheses*{S}}<2k^2$. Thus, choosing $\epsilon=\frac{1}{128 M_1 M_2 k^2}$ allows us to find $V^{*}$ up to an additive error of less than $\frac{1}{32}$ in time $\poly\parentheses*{k}$. We know that every information structure supported on $\Theta^n$ that achieves $V^{*}$ exactly maximizes $\mathbb{E}\brackets*{u_1}$ and maximizes $\mathbb{E}\brackets*{u_2}$ subject to maximizing $\mathbb{E}\brackets*{u_1}$, and thus maximizes the expectation of $u_3$ subject to the previous two optimizations. Since we can exactly compute in $\poly\parentheses*{k}$-time the values of the first two optimizations, we are able to compute $\frac{15}{16}e+\frac{\mc\parentheses*{V,E'}}{16}$ up to an additive error of less than $\frac{1}{32}$ in $\poly\parentheses*{k}$-time. Thus, we can compute $\mc\parentheses*{{V,E'}}$ in $\poly\parentheses*{k}$-time, implying $P=NP$.

It remains to show that it is possible to find $M_1,M_2$ as desired. The straightforward part is to show that for an appropriate choice of $M_2$, every sender-optimal information structure $\pi$ supported on $\Theta^n$ which is guaranteed to maximize $\mathbb{E}\brackets*{u_1\parentheses*{S}}$ must maximize $\mathbb{E}\brackets*{u_2\parentheses*{S}}$ subject to maximizing $\mathbb{E}\brackets*{u_1\parentheses*{S}}$. Indeed, suppose that such $\pi$ did not maximize $\mathbb{E}\brackets*{u_2\parentheses*{S}}$. From Step 3, the loss in $\mathbb{E}\brackets*{u_2\parentheses*{S}}$ compared to its tight upper bound is $\sum_{\{v,w\}\subseteq V:\; v\neq w}\frac{\mathbb{E}_{\pi}\brackets*{\parentheses*{X_v - X_w}^2}}{2}$.

It follows that with some probability $\rho>0$, not all the $X_v$s are equal. In this case, there are at least $k-1$ undirected pairs of $v,w$ with $X_v\ne X_w$. Since the $X_v$s take integer values, \emph{the loss in $\mathbb{E}\brackets*{u_2\parentheses*{S}}$ is at least $\rho\frac{k-1}{2}$}.

Let us bound the possible gain in $\mathbb{E}\brackets*{u_3\parentheses*{S}}$ compared to an information structure maximizing $\mathbb{E}\brackets*{u_2\parentheses*{S}}$ subject to maximizing $\mathbb{E}\brackets*{u_1\parentheses*{S}}$. First, note that always $u_3<2k^2$, and thus the total gain attributed to the event that not all the $X_v$s are equal is at most $\rho \cdot 2k^2$. Let us assume now that all the $X_v$s are equal. There are three cases:

\begin{itemize}
    \item If $X_v=0$ for every $v\in V$, then $u_3=0$.
    \item If $X_v=1$ for every $v\in V$, then $u_3\le MC\parentheses*{V,E'}$. Moreover, it follows from the probability computations in Step 4 (which apply verbatim, since we restrict attention to information structures maximizing $\mathbb{E}\brackets*{u_1\parentheses*{S}}$), that for every fixed $v^0\in V$, $\Pr[X_{v^0}=1 \wedge \omega=0]= 1/16$. Moreover, $\mathbb{E}\brackets*{u_3\parentheses*{S} \mid \forall v:\; X_v=1}\cdot \Pr[\forall v:\; X_v=1]=\mathbb{E}\brackets*{u_3\parentheses*{S} \mid \forall v:\; X_v=1,\omega=0}\cdot \Pr[\forall v:\; X_v=1,\omega=0]+\mathbb{E}\brackets*{u_3\parentheses*{S} \mid \forall v:\; X_v=1,\omega=1}\cdot \Pr[\forall v:\; X_v=1,\omega=1]$, and we know from the arguments in Step 4 that when $\omega=1$ and $X_v=1$ for every $v$, then $u_3=0$ (since Point 3 from Step 2 cannot hold). Therefore, $\mathbb{E}\brackets*{u_3\parentheses*{S} \mid \forall v:\; X_v=1}\cdot \Pr[\forall v:\; X_v=1]\le MC\parentheses*{V,E'}/16$.
    \item If $X_v=2$ for every $v\in V$, then $u_3 = 2e$. From the probability computations in Step 4, this event happens w.p.~at most $15/32$. Thus, $\mathbb{E}\brackets*{u_3\parentheses*{S} \mid \forall v:\; X_v=2}\cdot \Pr[\forall v:\; X_v=2]\le 15e/16$.
\end{itemize}

The computations above, together with the law of total expectation, imply that $\mathbb{E}\brackets*{u_3\parentheses*{S}}\le \frac{15}{16}e+\frac{\mc\parentheses*{V,E'}}{16} + \rho \cdot 2k^2$. Since an information structure maximizing $\mathbb{E}\brackets*{u_2\parentheses*{S}}$ subject to maximizing $\mathbb{E}\brackets*{u_1\parentheses*{S}}$ achieves $\mathbb{E}\brackets*{u_3\parentheses*{S}}= \frac{15}{16}e+\frac{\mc\parentheses*{V,E'}}{16}$, \emph{the potential gain from not maximizing $\mathbb{E}\brackets*{u_2\parentheses*{S}}$ is at most $\rho \cdot 2k^2$}. Thus, taking $M_2=4k^2$ ensures that the gain in $\mathbb{E}\brackets*{u_3\parentheses*{S}}$ is strictly less than the loss in $\mathbb{E}\brackets*{u_2\parentheses*{S}}$, a contradiction.

It remains to show that it is possible to choose a polynomially bounded $M_1$ for which every optimal information structure $\pi$ supported on $\Theta^n$ maximizes $\mathbb{E}\brackets*{u_1\parentheses*{S}}$. Let $\pi$ be some optimal information structure supported on $\Theta^n$ and suppose that it does not maximize  $\mathbb{E}\brackets*{u_1\parentheses*{S}}$. Let $v\in V$ be the index of some summand in the definition of $u_1$ whose expectation is not maximized by $\pi$, and denote the expectation of this summand by $E_v$.

Consider $\pi_{v}^0:=\pi|_{v_1,v_2,v_3} \mid \omega=0$ and   $\pi_{v}^1:=\pi|_{v_1,v_2,v_3} \mid \omega=1$, the information structures projected on the chain consisting of $v_1,v_2,v_3$ conditional on the state.\footnote{Here, the first $|$ means restriction, while the second $\mid$ means conditioning.} Let us replace those by $\tilde{\pi}_{v}^0,\tilde{\pi}_{v}^1$ that jointly maximize $E_v$ (we shall choose the specific $\tilde{\pi}_{v}^0,\tilde{\pi}_{v}^1$ among the maximizers later). It follows from Lemma~\ref{lem:lp} that they can be coupled with $\parentheses*{\pi_{w_i}^0,\pi_{w_i}^1}_{w\in V\setminus \{v\},\; i=1,2,3}$ into a joint information structure $\tilde{\pi}$ s.t. the joint distribution $\parentheses*{\pi_{w_i}^0,\pi_{w_i}^1}_{w\in V\setminus \{v\},\; i=1,2,3}$ remains unchanged. To finish the proof through contradiction, it is enough to show that an appropriate choice of $\tilde{\pi}$ is better for the sender than $\pi$ under an appropriate choice of $M_1$.

Note that maximizing $E_v$ can be done by maximizing a linear objective over an $O\parentheses*{1}$-sized bounded polyhedron, as follows from Lemma~\ref{lem:lp}. Let $C_v=O(1)$ be a global constant equal to the minimum nonzero difference between the values of $E_v$ at vertices of this polyhedron, and let $b_1,\ldots,b_l$ be these vertices. Assume w.l.o.g.~that $E_v$ is maximized at $b_1$, and let $N_v:=E_v\parentheses*{b_1}$. Let $B_{v}=\alpha_1 b_1+\ldots+\alpha_l b_l$ (for some $\alpha_1,\ldots,\alpha_l\ge 0$ s.t.~$\alpha_1+\ldots+\alpha_l=1$) be the point in $\co\parentheses*{b_1,\ldots,b_l}$ corresponding to $\pi_v$.

Let $\beta:=\sum_{j:\; E_v\parentheses*{b_j}<N_v} \alpha_j$. Then we have $N_v-E_v\parentheses*{B_v}\ge C_v \beta$, implying that transitioning from $\pi$ to $\tilde{\pi}$ improves $\mathbb{E}\brackets*{u_1}$ by at least $C_v \beta$ and by at most $N_v \beta$. Let $\tilde{B}_v:=\beta b_1 + \sum_{j:\; E_v\parentheses*{b_j}=N_v} \alpha_j b_j$. Note that $E_v\parentheses*{\tilde{B}_v}=N_v$. Take $\tilde{\pi}_{v}^0$ and $\tilde{\pi}_{v}^1$ to be the corresponding distributions over $\parentheses*{p_{v_1},p_{v_2},p_{v_3}}$ conditional on $\omega=0$ and $\omega=1$, respectively. Notably, the construction of $\tilde{\pi}$ from $\pi$ implies that w.p. at least~$\parentheses*{1-\beta}$, the drawn joint posterior over \emph{all} receivers remains unchanged.

In the definitions of $u_2,u_3$, the expectations of all the summands not containing $v$ are identical under $\pi$ and $\tilde{\pi}$. Note further that for $w\ne v$ and every information structure: $0\le X_vX_w\le 4$. Therefore, transitioning from $\pi$ to $\tilde{\pi}$ changes $\mathbb{E}\brackets*{u_2}$ by at most $4\parentheses*{k-1} \beta$.

Finally, note that for $\{v,w\}\in E'$ and every information structure:

\begin{align*}
    &0\le \mathbb{I}_{v_1,w_3\in S}+\mathbb{I}_{v_3,w_1\in S}\le 2.
\end{align*}

Hence, transitioning from $\pi$ to $\tilde{\pi}$ changes $\mathbb{E}\brackets*{u_3}$ by at most $2\parentheses*{k-1} \beta$.

To ensure that $\mathbb{E}_{\tilde{\pi}}\brackets*{u}>\mathbb{E}_{\pi}\brackets*{u}$, it is enough to pick $M_1$ for which $M_1 \cdot C_v \beta > M_2 \cdot 4\parentheses*{k-1} \beta + 2\parentheses*{k-1} \beta$. Picking $M_1:=\frac{8 M_2 k}{C_v}$ finishes the proof.

\end{proof}

\subsection{Approximate Tractability of the Supermodular Case}
\label{sub:super_easy}
In Subsection~\ref{sub:super}, we showed that even when $G$ is a disjoint union of chains of $3$ receivers, there is no FPTAS for the sender's optimal signaling scheme for a supermodular $u$ unless $P=NP$. However, this result does not rule out the existence of constant-factor approximation results. In this subsection, we show that it is possible to achieve a $\frac{1}{h+1}$-factor multiplicative approximation when $u$ is supermodular, where $h=h\parentheses*{G}$ is the depth of $G$.

\begin{proposition}
    \label{pro:super}
    For a monotone supermodular sender's utility $u$ and a forest communication hierarchy $G$ with depth $h$, a $\frac{1}{h+1}$ multiplicative approximation can be computed in time $\poly\parentheses*{n}$.
\end{proposition}

The proof idea is as follows. Let us pick an optimal \emph{straightforward} signaling scheme under private persuasion $\pi^{p}$, which can be computed in polynomial time by~\cite{arieli2019private} -- that is, under $\pi^{p}$, each receiver observes a private binary signal $s_i\in\{0,1\}$ which is interpreted as an incentive-compatible recommendation to take an action equal to the signal. From~\cite{arieli2019private}, $\pi^{p}$ sends $s_i=1$ to all receivers when $\omega=1$; conditional on $\omega=0$, $\pi^{p}$ draws some value $X\sim U\parentheses*{[0,1]}$, and sends $s_i=1$ to receiver $i$ iff $X\le \rho_i$ for a certain threshold $\rho_i$ which depends on $t_i$ and equals the maximum probability mass out of the prior probability for $\omega=0$ that can be pooled together with the prior probability for $\omega=1$ to make receiver $i$ weakly prefer $a_i=1$ over $a_i=0$. Specifically, $\rho_i:=\min\braces*{\frac{\mu\parentheses*{1-t_i}}{\parentheses*{1-\mu}t_i},1}$. Since the actual communication hierarchy $G$ allows a receiver to also observe the signals sent to the receivers he dominates, $\pi^{p}$ need no longer be incentive-compatible under $G$. The proof of Proposition~\ref{pro:super} is based on modifying $\pi^{p}$ to restore the incentive-compatibility property; it suffices since by~\cite{babichenko2022multi}, the sender's value under private persuasion is an upper bound for her value under any other communication hierarchy.

To this end, let us distinguish between the two possible states. Conditional on $\omega=0$, let us follow $\pi^{p}$  w.p.~$\frac{1}{h+1}$, and let us send the signal $0$ to all receivers w.p.~$1-\frac{1}{h+1}$. Intuitively, increasing the probability of sending the signal $0$ conditional on $\omega=0$ increases the persuasive power of observing a signal equal to $1$. Conditional on $\omega=1$, we shall carefully design a distribution over joint signals s.t.~it will follow $\pi^{p}$ w.p.~at least $\frac{1}{h+1}$. Moreover, we shall make sure that whenever some receiver $i$ directly receives the signal $1$, then regardless of the other signals that he observes, he will weakly prefer to take action $1$.

The idea of the construction of the distribution over joint signals conditional on $\omega=1$ is as follows. Let us denote by $s_i$ the signal sent directly to receiver $i$ while continuing to use $\sigma_i$ for the concatenation of all the signals observed by receiver $i$. Let us choose a depth $0\le l\le h$ uniformly at random. Note that the receivers of depth exactly $l$ do not dominate each other. For each receiver $i$ of depth $l$, we shall send $s_j=1$ to a subset of the subtree rooted at $i$ that follows $\pi^{p}$ conditional on $s_i=1$ \emph{and $\omega=0$ rather than $\omega=1$}. We shall further send  $s_j=1$ to all receivers with depth strictly less than $l$. In particular, w.p.~$\frac{1}{h+1}$ we shall choose $l=h$, which will lead to sending $s_i=1$ for every receiver $i$, thus following $\pi^{p}$. Moreover, following $\pi^{p}$ conditional on $s_i=1$ and $\omega=0$ with a high enough probability when $\omega=1$ serves the following purpose: Every concatenation of signals $\sigma_i$ that has $s_i=1$ and is observed when $\omega=0$ with a positive probability is also observed when $\omega=1$ with a probability large enough to make it incentive-compatible for receiver $i$ to pick action $a_i=1$.

\begin{proof}[Proof of Proposition~\ref{pro:super}]
Let $\pi^{p}$ be the optimal information structure under private persuasion provided by~\cite{arieli2019private}. Namely:

\begin{itemize}
\item Each receiver $i$ gets a binary signal $s_i\in\{0,1\}$ which is interpreted as a recommendation to take $a_i=s_i$, and all the recommendations are incentive compatible.
\item When $\omega=1$, $s_i=1$ for every $i$.
\item Define $\rho_i:=\min\braces*{\frac{\mu\parentheses*{1-t_i}}{\parentheses*{1-\mu}t_i},1}$ for $i=1,\ldots,n$. In other words, $\rho_i$ is the maximum possible probability of sending $s_i=1$ conditional on $\omega=0$ without violating incentive compatibility.
\item When $\omega=0$, draw $X\sim U\parentheses*{[0,1]}$ and set $s_i=1$ iff $X\le \rho_i$.
\end{itemize}

Let us define the information structure $\pi$ for the forest communication structure $G$ s.t. each receiver will get a binary signal $s_i\in\{0,1\}$ but will also observe the signals sent to all the receivers he dominates under $G$. $\pi$ is defined as follows:

When $\omega=0$, w.p.~$1-\frac{1}{h+1}$ send $s_i=0$ for all the receivers, and w.p.~$\frac{1}{h+1}$ follow $\pi^{p}$ conditional on $\omega=0$. When $\omega=1$, follow the procedure below:

\begin{itemize}
\item Choose a depth $l\in\braces*{0,\ldots,h}$ uniformly at random.
\item Send $s_i=1$ to all the receivers with depth strictly less than $l$.
\item For each receiver $i$ of depth exactly $l$, let $T_i$ be the subtree rooted at $i$. Define the distribution $\varphi_i$ over subsets $Q_i$ of $T_i$ containing $i$ by taking for every $Q\subseteq T_i$: $\Pr_{\varphi_i}[Q_i=Q]=\Pr_{\pi^{p}}[S\cap T_i=Q|\omega=0,\; i\in S]$.\footnote{If $\rho_i=0$, define $\varphi_i$ arbitrarily; for such a receiver $i$, getting $s_i=1$ implies $\omega=1$ w.p.~$1$.}
\item For each receiver $i$ of depth exactly $l$, draw $Q_i\sim \varphi_i$.
\item For a receiver $j$ of depth at least $l$, send $s_j=1$ if $j\in Q_i$ for the unique $i$ s.t. $j\in T_i$, and send $s_j=0$ otherwise.
\end{itemize}

Note that conditional on $\omega=0$, $\pi^{p}$ will be used w.p.~$\frac{1}{h+1}$. Moreover, conditional on $\omega=1$, w.p.~$\frac{1}{h+1}$ the procedure will choose $l=h$, which will result in all the receivers getting signal $1$, just as in $\pi^{p}$. Hence, in both states $\pi^{p}$ will be used w.p.~$\ge\frac{1}{h+1}$. Note that the sender's optimal value under private persuasion upper-bounds the sender's optimal value under any communication hierarchy~\cite{babichenko2022multi}. Thus, to finish the proof, it suffices to show that whenever a receiver will get the signal $1$, he will weakly prefer taking action $1$ regardless of the signals sent to the receivers he dominates. To this end, it suffices to show that for every $1\le i\le n$ and every possible value of $Q_i=\braces*{j\in D_i:\; s_j=1}\in\supp\parentheses*{\varphi_i}$, it holds that:

\begin{equation}
\label{eq:100}
  \Pr_{\pi}[\omega=1|Q_i]\ge \Pr_{\pi^{p}}[\omega=1|s_i=1], 
\end{equation}

or, equivalently:

\begin{equation}
\label{eq:500}
  \Pr_{\pi}[\omega=0|Q_i]\le \Pr_{\pi^{p}}[\omega=0|s_i=1]=\frac{\parentheses*{1-\mu}\rho_i}{\mu+\parentheses*{1-\mu}\rho_i}.
\end{equation}

Indeed, by the construction of $\pi$, we get that for any receiver $i$ and any possible value of $Q_i:=\braces*{j\in D_i:\; s_j=1}\in\supp\parentheses*{\varphi_i}$:\footnote{For simplicity of presentation, we identify $Q_i$ with the event that $Q_i$ takes a specific value.}

\begin{equation}
\label{eq:200}
  \Pr_{\pi}[\omega=1|Q_i]=\frac{\Pr_{\pi}[\omega=1\cap Q_i]}{\Pr_{\pi}[\omega=1\cap Q_i]+\Pr_{\pi}[\omega=0\cap Q_i]}.
\end{equation}

Note that for every $Q_i\in\supp\parentheses*{\varphi_i}$, we have $i\in Q_i$. Thus:

\begin{equation}
\label{eq:300}
  \Pr_{\pi}[\omega=0\cap Q_i]=\frac{1}{h+1}\Pr_{\pi^{p}}[\omega=0\cap Q_i]=\frac{1-\mu}{h+1}\Pr_{\pi^{p}}[Q_i|\omega=0].
\end{equation}

Moreover, by construction of $\pi$, we can lower-bound $\Pr_{\pi}[\omega=1\cap Q_i]$ with the probability of first choosing $l$ to be the depth of $i$ and then choosing $Q_i$. Namely, if $\rho_i>0$,

\begin{equation}
\label{eq:400}
\begin{aligned}
  &\Pr_{\pi}[\omega=1\cap Q_i]=\Pr[\omega=1]\Pr_{\pi}[Q_i|\omega=1]\ge\frac{1}{h+1}\mu\Pr_{\pi^{p}}[Q_i|\omega=0,s_i=1]\\
  &=\frac{\mu}{h+1}\frac{\Pr_{\pi^{p}}[Q_i\cap s_i=1|\omega=0]}{\Pr_{\pi^{p}}[s_i=1|\omega=0]}=\frac{\mu}{\parentheses*{h+1}\rho_i}\Pr_{\pi^{p}}[Q_i|\omega=0].
\end{aligned}
\end{equation}

Equations~(\ref{eq:200}),~(\ref{eq:300}), and~(\ref{eq:400}) imply:

\begin{align*}
  &\Pr_{\pi}[\omega=0|Q_i]=\frac{\Pr_{\pi}[\omega=0\cap Q_i]}{\Pr_{\pi}[\omega=1\cap Q_i]+\Pr_{\pi}[\omega=0\cap Q_i]}\\
  &\le\frac{\frac{1-\mu}{h+1}\Pr_{\pi^{p}}[Q_i|\omega=0]}{\frac{\mu}{\parentheses*{h+1}\rho_i}\Pr_{\pi^{p}}[Q_i|\omega=0]+\frac{1-\mu}{h+1}\Pr_{\pi^{p}}[Q_i|\omega=0]}=\frac{1-\mu}{\mu/\rho_i+\parentheses*{1-\mu}}=\frac{\parentheses*{1-\mu}\rho_i}{\mu+\parentheses*{1-\mu}\rho_i},
\end{align*}

which proves Equation~(\ref{eq:500}) and completes the proof.
\end{proof}

\subsection{Approximate Tractability of the Submodular Case}
\label{sub:sub}
As we know from~\cite{arieli2019private}, it is NP-hard to approximate the sender's value for submodular sender's utility up to a multiplicative factor larger than $\parentheses*{1-1/e}$ even for private persuasion. In this subsection, we show that for a forest communication structure with depth at most $h$, it is possible to achieve a multiplicative approximation ratio of $\frac{1}{h+1}\parentheses*{1-1/e}$ using a simple variation of the approximately optimal signaling scheme under private persuasion $\pi^{p}$. Namely, one should follow $\pi^{p}$ for receivers at a specific depth in $G$, chosen uniformly at random and revealed to all receivers, and reveal no additional information to other receivers (beyond the information that will be automatically revealed due to the structure of domination imposed by~$G$).

\begin{proposition}
    \label{pro:sub}
    For a monotone submodular sender's utility $u$ and any forest communication hierarchy $G$ with depth $h$, a $\frac{1}{h+1}\parentheses*{1-1/e}$ multiplicative approximation for an optimal signaling scheme can be computed in time $\poly\parentheses*{n}$.
\end{proposition}

\begin{proof}[Proof of Proposition~\ref{pro:sub}]
Let $\pi^{p}$ be an information structure achieving a $\parentheses*{1-1/e}$-factor approximation in polynomial time under private persuasion provided by~\cite{dughmi2017algorithmic2}. For $l=0,\ldots,h$, let $\pi^{p,l}$ be the signaling scheme that follows $\pi^{p}$ on the receivers of depth exactly $l$ and sends a totally uninformative signal directly to all the other receivers. Note that since receivers of the same depth cannot dominate each other, each $\pi^{p,l}$ indeed follows $\pi^p$ on receivers of depth $l$. We claim that $\pi$ obtained by picking $l\in\{0,\ldots,h\}$ uniformly at random, revealing $l$ to all receivers, and then following $\pi^{p,l}$, achieves a $\frac{1}{h+1}\parentheses*{1-1/e}$-factor~approximation.

Indeed, let $Q_l$ be the set of receivers of depth $l$ ($0\le l\le h$). We have by submodularity (which implies subadditivity) and monotonicity of $u$:

\begin{align*}
    &\mathbb{E}_{\pi^{p}}\brackets*{u\parentheses*{S}}\le\mathbb{E}_{\pi^{p}}\brackets*{\sum_{l=0}^{h} u\parentheses*{S\cap Q_l}}=\sum_{l=0}^{h}\mathbb{E}_{\pi^{p}}\brackets*{u\parentheses*{S\cap Q_l}}\le \sum_{l=0}^{h}\mathbb{E}_{\pi^{p,l}}\brackets*{u\parentheses*{S}}= \parentheses*{h+1}\mathbb{E}_{\pi}\brackets*{u\parentheses*{S}}.
\end{align*}

Since under any communication hierarchy the sender cannot achieve a higher value compared to private persuasion~\cite{babichenko2022multi}, the conclusion follows.
\end{proof}

\section{Tractability of the Symmetric Case}
\label{sec:sym}
In this section, we show that the problem of finding the optimal sender's value $V^{*}$ is computationally tractable when the sender's utility function $u$ is symmetric -- i.e., independent of the identities of the receivers who take action $1$ (Theorem~\ref{thm:chain}). The proof also provides a succinct representation for a signaling scheme that achieves $V^{*}$ (Remark~\ref{rem:sym}). However, it is not guaranteed that the support of such a signaling scheme has a polynomial size.

The latter issue suggests that the proof requires overcoming technical challenges. Indeed, we consider a relaxation of the LP from Lemma~\ref{lem:lp} that allows us to restrict attention, for every receiver $i$, only to his marginal posterior $p_i$ and the number of receivers he dominates that take action $1$, $k_i$. The relaxation will ensure that $\Pr[\omega=1 \mid p_i=q_i \wedge k_i=\kappa_i]=q_i$ and $\mathbb{E}\brackets*{p_j \mid p_i=q_i \wedge k_i=\kappa_i}=q_i$ for $i=1,\ldots,n$, $\kappa_i=0,\ldots,d_i-1$, $q_i\in\Theta$, $\parentheses*{j,i}\in E$. A recursive construction will show that the relaxation is actually tight -- i.e., every solution of the relaxed LP can be transformed into an inducible information structure with the same objective value.

The relaxation will still contain an exponential number of variables, since the variables will be the same as in the LP from Lemma~\ref{lem:lp}. Therefore, we shall look at the dual LP and provide a polynomial-time separation oracle. The separation oracle will use dynamic programming over the forest hierarchy $G$.

\begin{theorem}
    \label{thm:chain}
    Consider a forest communication hierarchy $G$. Suppose that the sender's utility function satisfies $u\parentheses*{S}=f\parentheses*{\absolute*{S}}$ for some monotone function $f:\brackets*{n}\cup \braces*{0} \to \mathbb{R}_{\ge 0}$ and every $S\subseteq\brackets*{n}$. Then there is a weakly polynomial-time algorithm computing $V^{*}$.
\end{theorem}

\begin{proof}[Proof of Theorem~\ref{thm:chain}]
Assume that $u\parentheses*{S}=f\parentheses*{\absolute*{S}}$. Let $E$ be the set of edges of $G$. For every receiver $1\le i\le n$ and $\bar{l}\in L$, let $k_i\parentheses*{\bar{l}}:=\absolute*{\{j\in D_i \setminus \{i\}:\; \tau_{l_j}\ge t_j\}}$.

Consider the following relaxation of LP~(\ref{lp:1}):

\begin{equation}
\label{lp:2}
\begin{aligned}
    \text{max}\;\;\;\;&\sum_{\bar{l}\in L} f\parentheses*{\absolute*{S\parentheses*{\bar{l}}}} \cdot \parentheses*{\mu x_{\bar{l}}^1+\parentheses*{1-\mu} x_{\bar{l}}^0}\\
    \text{s.t.}\;\;\;\;\;\;&\sum_{\bar{l}\in L:\; p\parentheses*{\bar{l}}_i=q_i,\; k_i\parentheses*{\bar{l}}=\kappa_i}\parentheses*{\parentheses*{1-q_i}\mu x_{\bar{l}}^{1} - q_i\parentheses*{1-\mu} x_{\bar{l}}^{0}}=0\;\;\;\;\;\;\;\;\; i=1,\ldots,n,\;\; \kappa_i=0,\ldots,d_i-1,\;\; q_i\in \Theta\\
    &\sum_{\bar{l}\in L:\; p\parentheses*{\bar{l}}_i=q_i,\; k_i\parentheses*{\bar{l}}=\kappa_i}\parentheses*{\tau_{l_j}-q_i}\parentheses*{\mu x_{\bar{l}}^{1} +\parentheses*{1-\mu} x_{\bar{l}}^{0}}=0\;\;\;\;\;\;\;\;\; \parentheses*{j,i}\in E,\;\; \kappa_i=0,\ldots,d_i-1,\;\; q_i\in \Theta\\
    &\sum_{\bar{l}\in L} x_{\bar{l}}^{w} = 1 \;\;\;\;\;\;\;\;\;\;\;\;\;\;\;\;\;\;\;\;\;\;\;\;\;\;\;\;\;\;\;\;\;\;\;\;\;\;\;\;\;\;\;\;\;\;\;\;\;\;\;\;\;\;\;\;\;\;\;\;\;\;\;\; w=0,1\\
    &x_{\bar{l}}^w\ge 0\;\;\;\;\;\;\;\;\;\;\;\;\;\;\;\;\;\;\;\;\;\;\;\;\;\;\;\;\;\;\;\;\;\;\;\;\;\;\;\;\;\;\;\;\;\;\;\;\;\;\;\;\;\;\;\;\;\;\;\;\;\;\;\;\;\;\;\;\;\; \bar{l}\in L,\;\; w=0,1.
    \end{aligned}
\end{equation}

The first family of constraints in LP~(\ref{lp:2}) is obtained by summing the constraints of the first family in LP~(\ref{lp:1}) for some fixed values of $i$, $k_i\parentheses*{\bar{l}}$, and $q_i$. Note that if $x_{\bar{l}}^w$ is interpreted as $\Pr[p=p\parentheses*{\bar{l}}\mid \omega=w]$, then $\sum_{\bar{l}\in L:\; p\parentheses*{\bar{l}}_i=q_i,\; k_i\parentheses*{\bar{l}}=\kappa_i} \parentheses*{\mu x_{\bar{l}}^1+\parentheses*{1-\mu} x_{\bar{l}}^0}$ is exactly $\Pr[p_i=q_i \wedge k_i=\kappa_i]$. Moreover, $\mu\sum_{\bar{l}\in L:\; p\parentheses*{\bar{l}}_i=q_i,\; k_i\parentheses*{\bar{l}}=\kappa_i} x_{\bar{l}}^1 = \Pr[\omega = 1 \wedge p_i=q_i \wedge k_i=\kappa_i]$. Thus, the first family of constraints ensures that $\Pr[\omega=1 \mid p_i=q_i \wedge k_i=\kappa_i]=q_i$.

The second family of constraints in LP~(\ref{lp:2}) is obtained by considering the difference of the following two quantities: (i)~the sum of all the constraints of the first family in LP~(\ref{lp:1}) for some fixed $i$, $q_i$, and $k_i\parentheses*{\bar{l}}=\kappa_i$; (ii)~the sum of all the constraints of the first family in LP~(\ref{lp:1}) for the parent $j$ of $i$ in $G$, $q_i$, and $k_i\parentheses*{\bar{l}}=\kappa_i$. The second family of constraints ensures that:

\begin{align*}
    &\mathbb{E}\brackets*{p_j \mid p_i=q_i \wedge k_i=\kappa_i}=\frac{1}{\Pr[p_i=q_i \wedge k_i=\kappa_i]}\sum_{\bar{l}\in L:\; p\parentheses*{\bar{l}}_i=q_i,\; k_i\parentheses*{\bar{l}}=\kappa_i} \tau_{l_j} \parentheses*{\mu x_{\bar{l}}^{1} +\parentheses*{1-\mu} x_{\bar{l}}^{0}}\\
    &=\frac{1}{\Pr[p_i=q_i \wedge k_i=\kappa_i]}\sum_{\bar{l}\in L:\; p\parentheses*{\bar{l}}_i=q_i,\; k_i\parentheses*{\bar{l}}=\kappa_i} q_i \parentheses*{\mu x_{\bar{l}}^{1} +\parentheses*{1-\mu} x_{\bar{l}}^{0}}=\frac{\Pr[\omega=1 \wedge p_i=q_i \wedge k_i=\kappa_i]}{\Pr[p_i=q_i \wedge k_i=\kappa_i]}\\
    &=\Pr[\omega=1 \mid p_i=q_i \wedge k_i=\kappa_i]=q_i.
\end{align*}

We shall prove that for every feasible solution of the relaxation LP~(\ref{lp:2}), there is an inducible information structure $\pi$ with sender's value equal to the value of the objective function for this solution. Let $X:=\parentheses*{x_{\bar{l}}^w}$ be a feasible solution of LP~(\ref{lp:2}). Note that it induces the two distributions $Y^{w}$ ($w=0,1$) over vectors of pairs $\parentheses*{q_i:=\tau_{l_i},\kappa_i}_{i=1}^{n}$ conditional on $\omega=w$. We know that $\Pr_{X}[\omega=1 \mid p_i=q_i \wedge k_i=\kappa_i]=q_i$ for every $q_i,\kappa_i$.

Let us traverse $G$ in preorder (i.e., from roots to the leaves, with an arbitrary ordering among the children of a node and also among the roots) and construct a distribution $X'$ over signals during the traversal. Given a state $\omega$, for the set $R$ of roots in $G$, the restriction of $X'$ to $R$ will draw $\parentheses*{q_i,\kappa_i}_{i\in R}$ according to $Y^{\omega}$ and send $\parentheses*{q_i,\kappa_i}$ directly to receiver $i$.  We shall make sure that according to $X'$:
\begin{enumerate}[label=\Roman*]
    \item $k_i=\kappa_i$ and $p_i=q_i$ for every $1\le i\le n$;
    \item the marginal posterior of every receiver $i$ depends on the signals of its descendants exclusively through $k_i$.
\end{enumerate}

Whenever we reach a certain node $j$, $X'$ will draw a vector $\parentheses*{q_i,\kappa_i}_{\parentheses*{j,i} \in E}$ according to probabilities specified by the projection of the distribution $\mu Y^{1}+\parentheses*{1-\mu} Y^{0} |\parentheses*{q_j,\kappa_j}$ on $\braces*{i:\; \parentheses*{j,i} \in E}$, and send $\parentheses*{q_i,\kappa_i}$ directly to receiver $i$. This will ensure that $\sum_{i:\; \parentheses*{j,i} \in E} \parentheses*{\kappa_i+\mathbb{I}_{q_i\ge t_i}} = \kappa_j$ w.p.~$1$.

By a trivial induction, we get $\Pr_{X'}[q_i,\kappa_i]=\Pr_{X}[p_i=q_i,k_i=\kappa_i]$ for every $i$. Moreover, by construction, for every $\parentheses*{j,i}\in E$, we have:

\begin{equation}\label{eq:ind1}
    \Pr_{X'}[q_i,\kappa_i \mid q_j,\kappa_j]=\Pr_{X}[p_i=q_i,k_i=\kappa_i \mid p_j=q_j,k_j=\kappa_j].
\end{equation}

Note further that our construction of $X'$ satisfies  condition (II). Indeed, consider some node $j$. Conditional on $\parentheses*{q_j,\kappa_j}$, the vector $\parentheses*{q_i,\kappa_i}_{\parentheses*{j,i} \in E}$ is independent of $\omega$ under $X'$. Moreover, the signals sent directly to all the descendants of each child $i$ of $j$ were also drawn independently of $\omega$ conditionally on $\parentheses*{q_i,\kappa_i}$, etc.

It remains to show that the posterior of each receiver $i$ is $q_i$ (the requirement $k_i=\kappa_i$ will then follow from the construction). We proceed be induction on the depth of $i$. Indeed, by construction, for every $i\in R$: $\Pr_{X'}[\omega=1 \mid q_i \wedge \kappa_i]=\Pr_{X}[\omega=1 \mid p_i=q_i \wedge k_i=\kappa_i]=q_i$. Given $i\notin R$, let $j$ be the parent of $i$, and assume that the condition holds for $j$. We get:

\begin{equation}\label{eq:ind2}
    \Pr_{X'}[\omega=1 | \parentheses*{q_i,\kappa_i}]=\mathbb{E}_{X'}\brackets*{\mathbb{I}_{\omega=1}\mid \parentheses*{q_i,\kappa_i}}=\mathbb{E}_{X'}\brackets*{\mathbb{E}_{X'}\brackets*{\mathbb{I}_{\omega=1}\mid \parentheses*{q_i,\kappa_i},\parentheses*{q_j,\kappa_j}}\mid \parentheses*{q_i,\kappa_i}}.
\end{equation}

Since $X'$ draws $\parentheses*{q_i,\kappa_i}_{\parentheses*{j,i} \in E}$, conditional on $\parentheses*{q_j,\kappa_j}$, independently of $\omega$, we get from the induction hypotheses:

\begin{equation}\label{eq:ind3}
    \mathbb{E}_{X'}\brackets*{\mathbb{I}_{\omega=1}\mid \parentheses*{q_i,\kappa_i},\parentheses*{q_j,\kappa_j}}=\mathbb{E}_{X'}\brackets*{\mathbb{I}_{\omega=1}\mid \parentheses*{q_j,\kappa_j}}=q_j.
\end{equation}

Thus:

\begin{align*}
&\Pr_{X'}[\omega=1 | \parentheses*{q_i,\kappa_i}]=_{(\ref{eq:ind2}),(\ref{eq:ind3})} \mathbb{E}_{X'}\brackets*{q_j \mid \parentheses*{q_i,\kappa_i}}=_{(\ref{eq:ind1})}\mathbb{E}_{X}\brackets*{p_j=q_j \mid \parentheses*{p_i,k_i}=\parentheses*{q_i,\kappa_i}}=q_i,
\end{align*}

where the last transition follows from $X$ being a feasible solution of LP~(\ref{lp:2}). Property (II) completes the induction.

Hence, our construction of $X'$ also implies (using the property $\sum_{i:\parentheses*{j,i}\in E} \parentheses*{k_i + \mathbb{I}_{p_i\ge t_i}}=k_j$) that $k_i=\kappa_i$ for every receiver $i$, which implies that the value of the objective is preserved by our construction. Thus, the values of the optimal solutions of LP~(\ref{lp:1}) and LP~(\ref{lp:2}) coincide, as needed.

The dual of LP~(\ref{lp:2}) is the following LP~(\ref{lp:3}):

\begin{equation}
\label{lp:3}
\begin{aligned}
    \text{min}\;\;\;\;&\alpha + \beta\\
    \text{s.t.}\;\;\;\;\;\;&\alpha + \mu\sum_{i=1}^{n} \parentheses*{1-\tau_{l_i}}\gamma^{i}_{k_i\parentheses*{\bar{l}},l_i} + \mu\sum_{\parentheses*{j,i}\in E} \parentheses*{\tau_{l_j} -\tau_{l_i}}\delta^{j,i}_{k_i\parentheses*{\bar{l}},l_i}\ge \mu f\parentheses*{\absolute*{S\parentheses*{\bar{l}}}}\;\;\;\;\;\;\;\;\;\;\;\;\;\;\;\;\;\;\;\;\;\;\;\;\;\;\;\;\;\;\;\;\;\;\;\;\; \bar{l}\in L\\
    &\beta - \parentheses*{1-\mu}\sum_{i=1}^{n} \tau_{l_i}\gamma^{i}_{k_i\parentheses*{\bar{l}},l_i}+\parentheses*{1-\mu}\sum_{\parentheses*{j,i}\in E} \parentheses*{\tau_{l_j} -\tau_{l_i}}\delta^{j,i}_{k_i\parentheses*{\bar{l}},l_i}\ge \parentheses*{1-\mu} f\parentheses*{\absolute*{S\parentheses*{\bar{l}}}}\;\;\;\;\;\;\;\;\;\;\;\;\;\;\;\;\; \bar{l}\in L,
    \end{aligned}
\end{equation}

where the dual variables are as follows:

\begin{itemize}
    \item $\alpha$ and $\beta$ correspond to the primal constraints $\sum_{\bar{l}\in L} x_{\bar{l}}^{w} = 1$ with $w=1$ and $w = 0$, respectively.
    \item $\gamma^{i}_{\kappa,l}$, $i=1,\ldots,n$, $\; 0\le \kappa<d_i$, $\; 0\le l\le m$ correspond to the first constraint family of LP~(\ref{lp:2}).
    \item $\delta^{j,i}_{\kappa,l}$, $\parentheses*{j,i}\in E$, $\; 0\le \kappa < d_i$, $\; 0\le l\le m$ correspond to the second constraint family of LP~(\ref{lp:2}).
\end{itemize}

 We shall provide a polynomial-time separation oracle for LP~(\ref{lp:3}). We shall focus on the first constraint family and vectors $\bar{l}\in L$ with $\absolute*{S\parentheses*{\bar{l}}}=k$ for a fixed $0\le k\le n$ (the treatment of the second family is analogous). Suppose that we are given the values of $\alpha,\beta$, and $\gamma^{i}_{\kappa,l}$, $\delta^{j,i}_{\kappa,l}$ for every $1\le i\le n$, $0\le \kappa< d_i$, $\parentheses*{j,i}\in E$, and $0\le l\le m$. It is enough to prove that we are able to find $\bar{l}\in L$ minimizing the function $\sum_{i=1}^{n} \parentheses*{1-\tau_{l_i}}\gamma^{i}_{k_i\parentheses*{\bar{l}},l_i} + \sum_{\parentheses*{j,i}\in E} \parentheses*{\tau_{l_j} -\tau_{l_i}}\delta^{j,i}_{k_i\parentheses*{\bar{l}},l_i}$ in polynomial time.

Indeed, let us write $j \prec i$ if receiver $i$ dominates receiver $j$. We shall use dynamic programming to compute a function $F\parentheses*{i,\kappa_i,l}$ ($i=1,\ldots,n,\; \kappa_i=0,\ldots,d_i-1,\; l=0,\ldots,m$) representing the minimum value of $\sum_{i'\preceq i} \parentheses*{1-\tau_{l_{i'}}}\gamma^{i'}_{k_{i'}\parentheses*{\bar{l}},l_{i'}} + \sum_{\parentheses*{j,i'}\in E:\; i'\prec i} \parentheses*{\tau_{l_j} -\tau_{l_{i'}}}\delta^{j,i'}_{k_{i'}\parentheses*{\bar{l}},l_{i'}}$ in polynomial time s.t.~$p_i=\tau_{l}$ and $k_i=\kappa_i$.

If $j$ is a leaf in $G$, then for $\kappa_j=0$ one can find $F\parentheses*{j,\kappa_j,l}$ by direct computation. Specifically, $F\parentheses*{j,\kappa_j=0,l}=\parentheses*{1-\tau_{l}}\gamma^{j}_{0,l}$. If $j$ is not a leaf, we can use the following recursive relation:

\begin{equation}\label{eq:rec}
    F\parentheses*{j,\kappa,l} = \parentheses*{1-\tau_{l}}\gamma^{j}_{\kappa,l} + \min_{\text{valid } \parentheses*{\kappa_i,l_i}_{i:\ \parentheses*{j,i}\in E}}\sum_{i:\ \parentheses*{j,i}\in E} \parentheses*{{F\parentheses*{i,\kappa_i,l_i} + \parentheses*{\tau_l - \tau_{l_i}} \delta^{j,i}_{\kappa_i,l_i}}},
\end{equation}

where the minimum is taken over all possible $\parentheses*{\kappa_i,l_i}_{i:\ \parentheses*{j,i}\in E}$ satisfying $\sum_{i} \parentheses*{\kappa_i + \mathbb{I}_{\tau_{l_i}\ge t_i}}=\kappa$. We shall explain why this minimum can be computed in polynomial time. Indeed, fix a non-leaf $j$ and $0\le l\le m$.

For a fixed child $i$ of $j$ and every $0 \le b \le d_i$, compute $\kappa_i\parentheses*{b}, l_i\parentheses*{b}$ that minimize the summand corresponding to $i$ in Equation~(\ref{eq:rec}) s.t. $\kappa_i\parentheses*{b} + \mathbb{I}_{\tau_{l_i\parentheses*{b}}\ge t_i} = b$; let $M_{i,b,l}$ be the minimum value of that summand -- that is, $M_{i,b,l}:=\min_{\kappa_i,l_i:\; \kappa_i + \mathbb{I}_{\tau_{l_i}\ge t_i} = b}\braces*{F\parentheses*{i,\kappa_i,l_i} + \parentheses*{\tau_l - \tau_{l_i}} \delta^{j,i}_{\kappa_i,l_i}}$. Let $i_1,\ldots, i_s$ be all the children of $j$, where $s=s\parentheses*{j} \ge 0$ is their total number. For $\xi=0,\ldots, s$ and $b = 0,\ldots, d_j - 1$, let us define $N^j_{\xi,b}$ by taking $N^j_{0,b}$ to be $0$ if $b=0$ and $+\infty$ if $b>0$, and for $\xi>0$ using the recursion $N^j_{\xi,b}=\min_{0\le \zeta \le \min\{b,d_{i_{\xi}}\}} \braces*{N^j_{\xi - 1, b - \zeta} + M_{i_{\xi},\zeta,l}}$. Note that this recursion allows us to compute $N^j_{s\parentheses*{j},b}$ in polynomial time for every fixed $j$ and $0\le b< d_j$. Moreover, we have for every fixed $j$ and $0 \le \kappa \le d_j - 1$: $$\min_{\text{valid } \parentheses*{\kappa_i,l_i}_{i:\ \parentheses*{j,i}\in E}}\sum_{i:\ \parentheses*{j,i}\in E} \parentheses*{{F\parentheses*{i,\kappa_i,l_i} + \parentheses*{\tau_l - \tau_{l_i}} \delta^{j,i}_{\kappa_i,l_i}}} = N^j_{s\parentheses*{j},\kappa}.$$

Thus, the recursion in Equation~(\ref{eq:rec}) can be computed in polynomial time. To finish the proof, one should perform one more step of the recursive process on an added dummy root $r$, with outgoing edges to all elements of $R$, s.t.~$r$ can never be persuaded to take action $1$ and has every $\gamma^r$ and $\delta^{r,i}$ equal to $0$. It remains just to check whether $\alpha+\mu \min_{0\le l\le m} F\parentheses*{r,k,l}\ge \mu f\parentheses*{k}$. If this condition is violated, one can find a violating $\bar{l}$ (one that minimizes $F$) using a standard backtracking~argument.
\end{proof}

\begin{remark}\label{rem:sym}
    Since our separation oracle provides a polynomial-time algorithm for the dual LP~(\ref{lp:3}), it is possible to recover an optimal solution $X$ of the primal LP~(\ref{lp:2}) in polynomial time. This solution yields a polynomial-time sampling procedure that allows us to draw signals from an optimal signaling scheme, since the distributions~$Y^{w}$ mentioned in Theorem~\ref{thm:chain} proof suffice for this purpose and they can be computed in polynomial time given $X$. However, it is not guaranteed that the support size of this optimal signaling scheme is polynomial. In particular, it might happen that there is no polynomial-time algorithm that provides an explicit representation for an optimal signaling scheme as a list of all possible joint posteriors coupled with the probabilities to induce them.
\end{remark}

\section{Computational Hardness of the Symmetric Case in the Non-Forest Setting}
\label{sec:nonforest}
In this section, we show that for a general directed graph $G$, unless $P=NP$, there is no FPTAS computing the optimal sender's value $V^{*}$ even if the sender's utility is a symmetric function whose range is $\{0,0.5,1\}$ and $G$ is a bipartite graph whose edges were directed from one side to the other.

\begin{proposition}
    \label{thm:nonforest1}
    Unless $P=NP$, there is no FPTAS for the problem of computing $V^{*}$ given a Bayesian persuasion instance with a communication hierarchy specified by an input directed graph $G$ all of whose directed paths have length of at most $1$ edge and a symmetric sender utility function that takes values only in $\braces*{0,0.5,1}$.
\end{proposition}

Similarly to the proof of~\cite{babichenko2022multi} that computing $V^{*}$ is NP-hard for a family of sender's utility functions they call \emph{separable supermajority functions}, we reduce from the decision version of the minimum $b$-union problem~\cite{vinterbo2002note}. However, our reduction uses a different construction and it further rules out the existence of an FPTAS unless $P=NP$. Specifically, suppose that we are given an instance of the minimum $b$-union problem: given are a base set $[l]$, $r\ge 2$ subsets $B_1,\ldots,B_r\subseteq [l]$, and positive integers $b,k$; we have to decide whether there are $b$ sets among $B_1,\ldots,B_r$ whose union is of size at most $k$. We construct a Bayesian persuasion instance as follows. Consider an instance with a uniform prior. Let us introduce a \emph{set receiver} $s_j$ for $1 \le j\le r$ that can only be persuaded to take action $1$ if he knows w.p.~$1$ that the state is $1$ (i.e., $t_{s_j}=1$); and an \emph{element receiver} $e_i$ for $1 \le i \le l$ with $t_{e_i}=0.5$. Let the edges of $G$ be all the pairs $\parentheses*{e_i,s_j}$ s.t.~$i\in B_j$. Then $G$ is a directed acyclic graph with all directed paths having a length of at most 1 edge. Finally, let $u\parentheses*{S}$ be $0$ if $\absolute*{S}<l-k$, $0.5$ if $l-k \le \absolute*{S} < l  + b$, and $1$ if $\absolute*{S}\ge l+b$.

The tradeoff between the sender's desire to exploit her ability to persuade the element receivers to take action $1$ ``for free" (i.e., without revealing any information) and the structural requirement to reveal the true state to a set receiver if the sender wants to persuade him to take action $1$ creates intricate dependencies that encode the  minimum $b$-union problem. Specifically, if there are $b$ sets with union size at most $k$, then revealing the true state exactly to the corresponding set receivers and the element receivers representing the elements in this union and revealing no information to other receivers achieves $V^{*}=\frac{3}{4}$. We further prove that if the union size for every $b$ sets is $>k$, then $V^{*}\le \frac{3}{4} - \frac{1}{2l^2k^2}$. For details, see Appendix~\ref{app:nonforest}.

\begin{remark}
    \label{rem:non_forest}
    The proof of Proposition~\ref{thm:nonforest1} also shows that there is no additive FPTAS for approximating $V^{*}$. Since the construction uses $u$ that takes values in $[0,1]$, this is a stronger statement.
\end{remark}

\section{Conclusions and Future Work}
\label{sec:conclusion}
We study the tractability of computing a sender-optimal signaling scheme in a multi-channel Bayesian persuasion setting with a forest communication hierarchy; each receiver has two possible actions, and the sender's utility is a function of the set of receivers taking a specific action. We extend the tractability results for symmetric and for modular sender's utility from the private persuasion framework of~\cite{babichenko2016computational,arieli2019private} to the forest hierarchy framework using a substantially different methodology. In contrast, we prove that the supermodular case does not admit an FPTAS (unless $P=NP$) even for forests of depth $2$. Yet, we provide an approximation algorithm for the supermodular case with the approximation ratio depending linearly on the depth of the forest, and an analogous algorithm for the submodular case. We further rule out an FPTAS, unless $P=NP$, for a non-forest hierarchy, even if the hierarchy is given by a bipartite graph whose edges are directed from one side to the other, and the sender's utility is a symmetric function that can only get three different~values.

Among the immediate open questions remaining, we should mention finding the tight approximation ratio for both the supermodular and the submodular cases. The possibility of extending our positive computational results to settings with more than two states is another interesting research direction. The question of generalizability of the results to frameworks beyond Bayesian persuasion, such as cheap talk~\cite{crawford1982strategic,babichenko2023algorithmic}, also remains~open. Given the leading example, it is also natural to consider a setting with multiple competing senders~\cite{gentzkow2016competition}. Finally, from a broader perspective, our methodology can enrich the toolbox available for analyzing algorithmic aspects of optimization of set functions. It will be interesting to generalize our techniques to provide both positive and negative algorithmic results in other subfields of computational economics and in combinatorial optimization in general.

\paragraph{Acknowledgements.} This work has been supported by the Center for Algorithms, Data, and Market Design at Yale (CADMY). Konstantin Zabarnyi is grateful to Yakov Babichenko and Benjamin Brooks for helpful discussions, as well as to his postdoc hosts, Dirk Bergemann and Yang~Cai.
\bibliographystyle{plainnat}
\bibliography{biblio}

@article{aumann2016agreeing,
  title={Agreeing to disagree},
  author={Aumann, Robert J and others},
  journal={Readings in Formal Epistemology},
  pages={859--862},
  year={2016},
  publisher={Springer}
}

@inproceedings{xu2015exploring,
  title={Exploring information asymmetry in two-stage security games},
  author={Xu, Haifeng and Rabinovich, Zinovi and Dughmi, Shaddin and Tambe, Milind},
  booktitle={Proceedings of the AAAI Conference on Artificial Intelligence},
  volume={29},
  number={1},
  year={2015}
}

@inproceedings{rabinovich2015information,
  title={Information disclosure as a means to security},
  author={Rabinovich, Zinovi and Jiang, Albert Xin and Jain, Manish and Xu, Haifeng},
  booktitle={Proceedings of the 2015 international conference on autonomous agents and multiagent systems},
  pages={645--653},
  year={2015}
}

@article{galperti2019belief,
  title={Belief meddling in social networks: An information-design approach},
  author={Galperti, Simone and Perego, Jacopo},
  journal={Available at SSRN 3340090},
  year={2019}
}

@article{galperti2020belief,
  title={Information Systems},
  author={Galperti, Simone and Perego, Jacopo},
  journal={Available at \url{https://www.jperego.com/files/papers/GP/draft.pdf}},
  year={2020}
}

@article{shin2024role,
  title={The role of messenger in advertising content: {B}ayesian persuasion perspective},
  author={Shin, Jiwoong and Wang, Chi-Ying},
  journal={Marketing Science},
  volume={43},
  number={4},
  pages={840--862},
  year={2024},
  publisher={INFORMS}
}

@article{kerman2025persuasion,
  title={Persuasion Gains and Losses from Peer Communication},
  author={Kerman, Toygar T and Tenev, Anastas P and Zabarnyi, Konstantin},
  journal={arXiv preprint arXiv:2509.09099},
  year={2025}
}

@article{babichenko2022regret,
  title={Regret-minimizing {B}ayesian persuasion},
  author={Babichenko, Yakov and Talgam-Cohen, Inbal and Xu, Haifeng and Zabarnyi, Konstantin},
  journal={Games and Economic Behavior},
  volume={136},
  pages={226--248},
  year={2022},
  publisher={Elsevier}
}

@article{sherali1982equivalent,
  title={Equivalent weights for lexicographic multi-objective programs: Characterizations and computations},
  author={Sherali, Hanif D},
  journal={European Journal of Operational Research},
  volume={11},
  number={4},
  pages={367--379},
  year={1982},
  publisher={Elsevier}
}

@inproceedings{dughmi2016algorithmic,
  title={Algorithmic {B}ayesian persuasion},
  author={Dughmi, Shaddin and Xu, Haifeng},
  booktitle={Proceedings of the forty-eighth annual ACM symposium on Theory of Computing},
  pages={412--425},
  year={2016}
}

@article{behringer1977lexicographic,
  title={Lexicographic quasiconcave multiobjective programming},
  author={Behringer, Fred Alois},
  journal={Zeitschrift f{\"u}r Operations Research},
  volume={21},
  number={3},
  pages={103--116},
  year={1977},
  publisher={Springer}
}

@article{crawford1982strategic,
  title={Strategic information transmission},
  author={Crawford, Vincent P and Sobel, Joel},
  journal={Econometrica: Journal of the Econometric Society},
  pages={1431--1451},
  year={1982},
  publisher={JSTOR}
}

@article{isermann1982linear,
  title={Linear lexicographic optimization},
  author={Isermann, H},
  journal={Operations-Research-Spektrum},
  volume={4},
  number={4},
  pages={223--228},
  year={1982},
  publisher={Springer}
}

@book{schrijver1998theory,
  title={Theory of linear and integer programming},
  author={Schrijver, Alexander},
  year={1998},
  publisher={John Wiley \& Sons}
}

@inproceedings{karp1972reducibility,
  title={Reducibility among combinatorial problems},
  author={Karp, Richard M},
  booktitle={Complexity of Computer Computations: Proceedings of a symposium on the Complexity of Computer Computations, held March 20--22, 1972, at the IBM Thomas J. Watson Research Center, Yorktown Heights, New York, and sponsored by the Office of Naval Research, Mathematics Program, IBM World Trade Corporation, and the IBM Research Mathematical Sciences Department},
  pages={85--103},
  year={1972},
  organization={Springer}
}

@inproceedings{arieli2022persuasion,
  title={Persuasion as Transportation.},
  author={Arieli, Itai and Babichenko, Yakov and Sandomirskiy, Fedor},
  booktitle={EC},
  pages={468},
  year={2022}
}

@article{babichenko2023algorithmic,
  title={Algorithmic cheap talk},
  author={Babichenko, Yakov and Talgam-Cohen, Inbal and Xu, Haifeng and Zabarnyi, Konstantin},
  journal={arXiv preprint arXiv:2311.09011},
  year={2023}
}

@article{Candogan20,
	author    = {Candogan, Ozan},
	title     = {Information design in operations},
	journal   = {TutORials in Operations Research},
	pages     = {176--201},
	year      = {2020}
}

@article{dughmi2017algorithmic,
  title={Algorithmic information structure design: a survey},
  author={Dughmi, Shaddin},
  journal={ACM SIGecom Exchanges},
  volume={15},
  number={2},
  pages={2--24},
  year={2017},
  publisher={ACM New York, NY, USA}
}

@article{bergemann2019information,
  title={Information design: A unified perspective},
  author={Bergemann, Dirk and Morris, Stephen},
  journal={Journal of Economic Literature},
  volume={57},
  number={1},
  pages={44--95},
  year={2019}
}

@inproceedings{babichenko2022multi,
  title={Multi-Channel {B}ayesian Persuasion},
  author={Babichenko, Yakov and Talgam-Cohen, Inbal and Xu, Haifeng and Zabarnyi, Konstantin},
  booktitle={13th Innovations in Theoretical Computer Science Conference (ITCS 2022)},
  year={2022},
  organization={Schloss Dagstuhl-Leibniz-Zentrum f{\"u}r Informatik}
}

@article{mathevet2020information,
  title={On information design in games},
  author={Mathevet, Laurent and Perego, Jacopo and Taneva, Ina},
  journal={Journal of Political Economy},
  volume={128},
  number={4},
  pages={1370--1404},
  year={2020},
  publisher={The University of Chicago Press Chicago, IL}
}

@inproceedings{lingenbrink2018signaling,
  title={Signaling in online retail: Efficacy of public signals},
  author={Lingenbrink, David and Iyer, Krishnamurthy},
  booktitle={Proceedings of the 13th Workshop on Economics of Networks, Systems and Computation},
  pages={1--1},
  year={2018}
}

@article{che2018recommender,
  title={Recommender systems as mechanisms for social learning},
  author={Che, Yeon-Koo and H{\"o}rner, Johannes},
  journal={The Quarterly Journal of Economics},
  volume={133},
  number={2},
  pages={871--925},
  year={2018},
  publisher={Oxford University Press}
}

@article{kuccukgul2022engineering,
  title={Engineering social learning: Information design of time-locked sales campaigns for online platforms},
  author={K{\"u}{\c{c}}{\"u}kg{\"u}l, Can and {\"O}zer, {\"O}zalp and Wang, Shouqiang},
  journal={Management Science},
  volume={68},
  number={7},
  pages={4899--4918},
  year={2022},
  publisher={INFORMS}
}

@article{du2025optimal,
  title={Optimal Calibrated Signaling in Digital Auctions},
  author={Du, Zhicheng and Tang, Wei and Wang, Zihe and Zhang, Shuo},
  journal={arXiv preprint arXiv:2507.17187},
  year={2025}
}

@article{kolotilin2025persuasion,
  title={Persuasion and matching: Optimal productive transport},
  author={Kolotilin, Anton and Corrao, Roberto and Wolitzky, Alexander},
  journal={Journal of Political Economy},
  volume={133},
  number={4},
  pages={1334--1381},
  year={2025},
  publisher={The University of Chicago Press Chicago, IL}
}

@article{dworczak2024persuasion,
  title={The persuasion duality},
  author={Dworczak, Piotr and Kolotilin, Anton},
  journal={Theoretical Economics},
  volume={19},
  number={4},
  pages={1701--1755},
  year={2024},
  publisher={Wiley Online Library}
}

@article{mahzoon2022hierarchical,
  title={Hierarchical {B}ayesian persuasion: Importance of vice presidents},
  author={Mahzoon, Majid},
  journal={arXiv preprint arXiv:2204.05304},
  year={2022}
}

@article{mathevet2020organized,
  title={Organized Information Transmission},
  author={Mathevet, Laurent and Taneva, Ina},
  journal={Available at SSRN 3656555},
  year={2020}
}

@article{kamenica2019bayesian,
  title={{B}ayesian persuasion and information design},
  author={Kamenica, Emir},
  journal={Annual Review of Economics},
  volume={11},
  pages={249--272},
  year={2019},
  publisher={Annual Reviews}
}

@inproceedings{bro2012send,
  title={Send mixed signals: earn more, work less},
  author={Bro Miltersen, Peter and Sheffet, Or},
  booktitle={Proceedings of the 13th ACM Conference on Electronic Commerce},
  pages={234--247},
  year={2012}
}

@article{bergemann2016bayes,
  title={Bayes correlated equilibrium and the comparison of information structures in games},
  author={Bergemann, Dirk and Morris, Stephen},
  journal={Theoretical Economics},
  volume={11},
  number={2},
  pages={487--522},
  year={2016},
  publisher={Wiley Online Library}
}

@article{kamenica2011bayesian,
  author = {Emir Kamenica and Matthew Gentzkow},
  title = {{B}ayesian persuasion},
  journal = "American Economic Review",
  volume = "101",
  number = "6",
  pages = "2590--2615",
  year = "2011"
}

@article{vinterbo2002note,
  title={A note on the hardness of the k-ambiguity problem},
  author={Vinterbo, Staal A.},
  journal={Harvard Med. School, Boston, MA, USA, Tech. Rep. DSG},
  year={2002},
  publisher={Citeseer}
}

@article{arieli2019private,
  author = {Itai Arieli and Yakov Babichenko},
  title = {Private {B}ayesian persuasion},
  journal = "Journal of Economic Theory",
  volume = "182",
  pages = "185--217",
  year = "2019",
  publisher="Elsevier"
}

@article{gentzkow2016competition,
  title={Competition in persuasion},
  author={Gentzkow, Matthew and Kamenica, Emir},
  journal={The Review of Economic Studies},
  volume={84},
  number={1},
  pages={300--322},
  year={2016},
  publisher={Review of Economic Studies Ltd}
}

@book{horngren2009cost,
  title={Cost accounting: A managerial emphasis},
  author={Horngren, Charles T},
  year={2009},
  publisher={Pearson Education India}
}

@article{Blackwell53,
  title={Equivalent comparisons of experiments},
  author={David Blackwell},
  journal={The Annals of Mathematical Statistics},
  pages={265--272},
  year={1953},
  publisher={JSTOR}
}

@book{aumann1995repeated,
  author = {Robert J.~Aumann and Michael Maschler and Richard E.~Stearns},
  title = "Repeated Games with Incomplete Information",
  year = "1995",
  publisher = "MIT Press"
}

@book{shaked2007stochastic,
  title={Stochastic orders},
  author={Shaked, Moshe and Shanthikumar, J George},
  year={2007},
  publisher={Springer}
}

@article{drakopoulos2021persuading,
  title={Persuading customers to buy early: The value of personalized information provisioning},
  author={Drakopoulos, Kimon and Jain, Shobhit and Randhawa, Ramandeep},
  journal={Management Science},
  volume={67},
  number={2},
  pages={828--853},
  year={2021},
  publisher={INFORMS}
}

@inproceedings{dughmi2017algorithmic2,
  author = {Shaddin Dughmi and Haifeng Xu},
  title = {Algorithmic persuasion with no externalities},
  booktitle = "Proceedings of the 2017 {ACM} Conference on Economics and Computation {EC}",
  pages = "351--368",
  year = "2017"
}

@article{candogan2022persuasion,
  title={Persuasion in networks: Public signals and cores},
  author={Candogan, Ozan},
  journal={Operations Research},
  volume={70},
  number={4},
  pages={2264--2298},
  year={2022},
  publisher={INFORMS}
}

@article{haghtalab2024leakage,
  title={Leakage-robust Bayesian persuasion},
  author={Haghtalab, Nika and Qiao, Mingda and Yang, Kunhe},
  journal={arXiv preprint arXiv:2411.16624},
  year={2024}
}

@inproceedings{castiglioni2020persuading,
  title={Persuading Voters: It's Easy to Whisper, It's Hard to Speak Loud},
  author={Castiglioni, Matteo and Celli, Andrea and Gatti, Nicola},
  booktitle={Proceedings of the AAAI Conference on Artificial Intelligence},
  volume={34},
  number={02},
  pages={1870--1877},
  year={2020}
}

@inproceedings{castiglioni2021persuading,
  title={Persuading Voters in District-based Elections},
  author={Castiglioni, Matteo and Gatti, Nicola},
  booktitle={Proceedings of the AAAI Conference on Artificial Intelligence},
  volume={35},
  number={6},
  pages={5244--5251},
  year={2021}
}

@techreport{blackwell1950comparison,
  title={Comparison of experiments},
  author={Blackwell, David},
  year={1950},
  institution={Howard University, Washington, United States}
}

@article{alonso2016persuading,
  title={Persuading voters},
  author={Ricardo Alonso and Odilon C{\^a}mara},
  journal={American Economic Review},
  volume={106},
  number={11},
  pages={3590--3605},
  year={2016}
}

@article{candogan2020information,
  title={On information design with spillovers},
  author={Candogan, Ozan and Guo, Yifan and Xu, Haifeng},
  journal={Available at SSRN 3537289},
  year={2020}
}

@misc{Mean,
    author = "{Wikipedia contributors}",
    title = "Mean-preserving spread
 -- {Wikipedia}{,} The Free Encyclopedia",
    year = "2026",
    url = "https://en.wikipedia.org/wiki/Mean-preserving_spread",
    note = "[Online; accessed 26-July-2026]"
  }

@article{babichenko2016computational,
  title={Computational aspects of private {B}ayesian persuasion},
  author={Babichenko, Yakov and Barman, Siddharth},
  journal={arXiv preprint arXiv:1603.01444},
  year={2016}
}

@article{arieli2021feasible,
  title={Feasible joint posterior beliefs},
  author={Arieli, Itai and Babichenko, Yakov and Sandomirskiy, Fedor and Tamuz, Omer},
  journal={Journal of Political Economy},
  volume={129},
  number={9},
  pages={2546--2594},
  year={2021},
  publisher={The University of Chicago Press Chicago, IL}
}

@inproceedings{badanidiyuru2018targeting,
  author = "Badanidiyuru, Ashwinkumar and Bhawalkar, Kshipra and Xu, Haifeng",
  title = "Targeting and signaling in ad auctions",
  booktitle = "Proceedings of the 29th Annual ACM-SIAM Symposium on Discrete Algorithms",
  pages = "2545--2563",
  year = "2018",
  organization="SIAM"
}

@article{wang2013bayesian,
  title={{B}ayesian persuasion with multiple receivers},
  author={Wang, Yun},
  journal={Available at SSRN 2625399},
  year={2013}
}

@inproceedings{dughmi2014hardness,
  title={On the hardness of signaling},
  author={Dughmi, Shaddin},
  booktitle={2014 IEEE 55th Annual Symposium on Foundations of Computer Science},
  pages={354--363},
  year={2014},
  organization={IEEE}
}

@inproceedings{cheng2015mixture,
  title={Mixture selection, mechanism design, and signaling},
  author={Yu Cheng and Ho Yee Cheung and Shaddin Dughmi and Ehsan Emamjomeh-Zadeh and Li Han and Shang-Hua Teng},
  booktitle={IEEE 56th Annual Symposium on Foundations of Computer Science (FOCS 2015)},
  year={2015},
}

@inproceedings{xu2020tractability,
  title={On the tractability of public persuasion with no externalities},
  author={Xu, Haifeng},
  booktitle={Proceedings of the Fourteenth Annual ACM-SIAM Symposium on Discrete Algorithms},
  pages={2708--2727},
  year={2020},
  organization={SIAM}
}

@article{brooks2022information,
  title={Information hierarchies},
  author={Brooks, Benjamin and Frankel, Alexander and Kamenica, Emir},
  journal={Econometrica},
  volume={90},
  number={5},
  pages={2187--2214},
  year={2022},
  publisher={Wiley Online Library}
}

\appendix
\section{Proof of Lemma~\ref{lem:grid}}
\label{app:prelim}
\begin{proof}[Proof of Lemma~\ref{lem:grid}]
    Fix an information structure $\pi$ induced by some signaling scheme $\Sigma$. With a slight abuse of notation, define $\supp\parentheses*{\pi}:=\cup_{i=1}^n \supp\parentheses*{\pi_i}$. Suppose that $\supp\parentheses*{\pi}\setminus\Theta\neq\emptyset$. Take some $i\in\brackets*{n}$, $j\in\brackets*{m}$ s.t.~$\Pr_{p_i\sim\pi_i}[p_i\in\parentheses*{\tau_{j-1},\tau_j}]>0$, and all the ancestors of $i$ in $G$ have marginal posteriors already discretized to $\Theta$.
    
    Let us condition on receiver $i$ observing a certain value $s$ of the concatenated signal $\sigma_i$ s.t.~the corresponding marginal posterior of receiver $i$ is some $q\in \parentheses*{\tau_{j-1},\tau_j}$. Let $\alpha:=\Pr[\sigma_i=s]$. There is a unique way to move the probability mass $\alpha$ on $q$ to $\tau_{j-1}$ and $\tau_j$ in a mean-preserving way -- i.e., the mass of $\alpha$ on $q$ is replaced by a mass of $\beta\alpha$ on $\tau_{j-1}$ and $\parentheses*{1-\beta}\alpha$ on $\tau_j$, with $\beta \tau_{j-1}+\parentheses*{1-\beta}\tau_j=q$.\footnote{A similar transformation can also be performed when the distribution $\pi_i$ is not discrete; for simplicity of presentation, we focus on discrete distributions here.} If $i$ is a root, then the transformation can be done without violating the inducibility of the information structure. Otherwise, let $\bar{i}$ be the parent of $i$ in $G$. We first claim that the resultant distribution $\nu$ over $\braces*{\tau_{j-1},\tau_j}$ is a mean-preserving contraction of $\pi_{\bar{i}}|\parentheses*{\sigma_i=s}$.
    
    Suppose that the probability mass $\alpha>0$ on $q$ was originally split in $\pi_{\bar{i}}$ to masses $\gamma_l$ on the respective $\tau_l$s ($0\le l\le m$) -- i.e., $\sum_l \gamma_l \tau_l = \alpha q$ and $\sum_l \gamma_l = \alpha$. To show that $\pi_{\bar{i}}|\parentheses*{\sigma_i=s}$ is a mean-preserving spread of $\nu$, we have to show that there exist nonnegative $x_l$s and $y_l$s ($0\le l\le m$) with $\sum_l x_l = \beta\alpha$, $\sum_l y_l = \parentheses*{1-\beta}\alpha$, $\sum_l x_l \tau_l = \beta\alpha \tau_{j-1}$, $\sum_l y_l \tau_l = \parentheses*{1-\beta}\alpha \tau_j$, and $x_l+y_l=\gamma_l$ for $l=0,\ldots,m$; note that the second and the fourth constraints follow from the other constraints, and thus can be ignored.
    
    We only have to show that the following system of two constraints has a solution, with $0\le x_l \le \gamma_l$ for every $l$: $\sum_l x_l = \beta\alpha$ and $\sum_l x_l \tau_l = \beta\alpha \tau_{j-1}$. Note that setting $x_l=\beta\gamma_l$ for every $l$ satisfies the first constraint, but we get $\sum_l x_l \tau_l = \beta\alpha q > \beta\alpha \tau_{j-1}$. Let $s$ be the smallest integer s.t.~$\sum_{l\le s} \gamma_l \ge \beta\alpha$. If $\sum_{l < s} \gamma_l \tau_l + \parentheses*{\beta\alpha - \sum_{l < s} \gamma_l} \tau_s \le \beta\alpha \tau_{j-1}$, then we can make sure that the first constraint holds for some values of $x_l$s with $x_l=0$ for any $l>s$, implying  $\sum_l x_l \tau_l \le \beta\alpha \tau_{j-1}$, and a convexity argument shows that there is a solution simultaneously to both constraints, with $0\le x_l\le \gamma_l$ for every $l$.

    Assume now $\sum_{l < s} \gamma_l \tau_l + \parentheses*{\beta\alpha - \sum_{l < s} \gamma_l} \tau_s > \beta\alpha \tau_{j-1}$. In particular, we get $j\le s$ (since $\tau_1\le \tau_2\le\ldots\le \tau_{m}$). Moreover, we must have $\sum_{l > s} \gamma_l \tau_l + \parentheses*{\sum_{l \le s} \gamma_l - \beta\alpha} \tau_s \ge \parentheses*{\sum_{l > s} \gamma_l + \parentheses*{\sum_{l \le s} \gamma_l - \beta\alpha}} \tau_s=\parentheses*{1-\beta}\alpha \tau_s \ge\parentheses*{1-\beta}\alpha \tau_j$. Summing up the last two inequalities yields $\sum_l \gamma_l \tau_l > \beta\alpha \tau_{j-1}+\parentheses*{1-\beta}\alpha \tau_j=\alpha q$, a contradiction. Thus, indeed $\pi_{\bar{i}}|\parentheses*{\sigma_i=s}$ is a mean-preserving spread of $\nu$.
    
   Let $Y\sim \pi_{\bar{i}}$. Let us define the random variable $X$ as follows. Conditional on $\sigma_i = s$ and $Y=y$ for some $s\in\supp\parentheses*{\sigma_i}$ and $y\in\supp\parentheses*{Y}$, $X$ will follow the following distribution using a fresh random drawing: $\Pr[X=\tau_{j-1}\mid \sigma_i = s,\; Y=y]=\frac{x_l}{\gamma_l}$ and $\Pr[X=\tau_{j}\mid \sigma_i = s,\; Y=y]=\frac{y_l}{\gamma_l}$. In particular, conditional on $\sigma_i = s$, $X$ exactly follows $\nu$. Moreover, $\mathbb{E}\brackets*{Y\mid X, \sigma_i = s} = X$.
    
    Create a signaling scheme $\Sigma_s$ from $\Sigma$ by, conditional on $\sigma_i=s$ under $\Sigma$, sending $X$ directly to receiver $i$ in addition to the information that he already observes. Then receiver $\bar{i}$ and all his ancestors do not get additional information from observing $X$. Moreover, since $\mathbb{E}\brackets*{Y\mid X, \sigma_i = s} = X$, the marginal posterior $p_i$ under $\Sigma_s$ conditional on $\sigma_i=s$ under $\Sigma$ follows $\nu$. Finally, all receivers who do not dominate $i$ and are different from $i$ get exactly the same information as under $\Sigma$. Thus, the information structure induced by $\Sigma_s$ is identical to $\pi$ on all marginals except for $\pi_i$, for which $q$ is spread according to $\nu$. The expected sender's utility cannot decrease, since we only moved probability mass from $\parentheses*{\tau_{j-1},\tau_j}$ to its endpoints, and thus no receiver who took action $1$ before the transformation could take action $0$ after it.
    
    Note that for a given receiver $i$ and an interval $\parentheses*{\tau_{j-1},\tau_j}$ with $\Pr[p_i\in\parentheses*{\tau_{j-1},\tau_j}]>0$, the above transformation can be performed simultaneously for all values $s$ of $\sigma_i$ s.t. ~the corresponding marginal posterior of receiver $i$ is in $\parentheses*{\tau_{j-1},\tau_j}$. Repeating this procedure as long as there are receivers with $\supp\parentheses*{\pi}\setminus\Theta\neq\emptyset$ completes the proof.
\end{proof}

\section{Proof of Proposition~\ref{pro:linear}}
\label{app:modular}
\begin{proof}[Proof of Proposition~\ref{pro:linear}]

    Consider the following LP~(\ref{lp:0}):
    
\begin{equation}
\label{lp:0}
\begin{aligned}
    \text{max}\;\;\;\;&\sum_{i=1}^{n} \lambda_i \cdot \sum_{l: \tau_l\ge t_i} x_{i,l}\\
    \text{s.t.}\;\;\;\;\;\;&\sum_{l=0}^{m} \tau_{l} x_{i,l}=\mu\;\;\;\;\;\;\;\;\;\;\;\;\;\;\;\;\;\;\;\;\;\;\;\;\;\;\;\;\;\;\;\;\;\;\;\;\;\;\;\;\;\;\;\;\;\;\;\;\;\;\;\;\;\;\; i=1,\ldots,n\\
    &\sum_{l=0}^{l_0} x_{j,l} \cdot \parentheses*{\tau_{l_0}-\tau_{l}} \le \sum_{l=0}^{l_0} x_{i,l} \cdot \parentheses*{\tau_{l_0}-\tau_{l}} \;\;\;\;\;\;\;\;\;\;\;\;\;\;\; \parentheses*{i,j}\in E,\; l_0=0,\ldots, m\\
    &\sum_{l=0}^{m} x_{i,l} = 1 \;\;\;\;\;\;\;\;\;\;\;\;\;\;\;\;\;\;\;\;\;\;\;\;\;\;\;\;\;\;\;\;\;\;\;\;\;\;\;\;\;\;\;\;\;\;\;\;\;\;\;\;\;\;\;\;\;\;\; i=1,\ldots,n\\
    &x_{i,l}\ge 0\;\;\;\;\;\;\;\;\;\;\;\;\;\;\;\;\;\;\;\;\;\;\;\;\;\;\;\;\;\;\;\;\;\;\;\;\;\;\;\;\;\;\;\;\;\;\;\;\;\;\;\;\;\;\;\;\;\;\;\;\;\;\;\; i=1,\ldots,n,\;l=0,\ldots,m.
    \end{aligned}
\end{equation}

The variables are $x_{i,l}$, $\; i=1,\ldots, n$, $\; l=0,\ldots, m$, where $x_{i,l}$ is interpreted as $\Pr[p_i=\tau_l]$. Since the numbers of variables and constraints are $\poly\parentheses*{n}$, and each $\lambda_i$ can be found by a single query to $u$, LP~(\ref{lp:0}) can be solved in polynomial time for every given $u$. By Lemma~\ref{lem:grid}, when aiming to maximize the sender's expected utility, restricting attention to information structures supported on $\Theta^n$ is w.l.o.g. We shall show that there is a one-to-one correspondence between marginals $\pi_1,\ldots,\pi_n$ of inducible information structures $\pi$ supported on $\Theta^n$ and feasible solutions of LP~(\ref{lp:0}) s.t.~the sender's value $V\parentheses*{\pi}$ equals the value of the objective function in LP~(\ref{lp:0}).\footnote{Note that since $u$ is modular, the marginals $\pi_1,\ldots,\pi_n$ suffice to compute $V\parentheses*{\pi}$. However, there may be multiple ways to couple the marginals into a joint information structure.}

Fix an inducible information structure $\pi$ supported on $\Theta^n$. Define the variables $x_{i,l}=\Pr_{\pi}[p_i=\tau_l]$, $\; i=1,\ldots, n$, $\; l=0,\ldots, m$.  Then the objective function indeed equals the expected sender's utility over $\pi$. The first family of constraints is satisfied due to Bayes-plausibility, while the second family of constraints holds since the distribution over the marginal posterior $p_i$ is a mean-preserving spread of the distribution over $p_j$ whenever $\parentheses*{i,j}\in E$.\footnote{See, e.g., Theorem 4.A.2 in~\cite{shaked2007stochastic}; the simple criterion for the discrete case is immediately obtained using the tail sum formula for expectation.} The last two families hold since $\parentheses*{x_{i,l}}_{0\le l\le m}$ is a probability distribution for a fixed $i$. We showed that every information structure defines a feasible solution of LP~(\ref{lp:0}) with the value of the corresponding objective function being the expected sender's utility.

Conversely, take a feasible solution $\parentheses*{x_{i,l}}_{i=1,\ldots, n, l=0,\ldots, m}$ of LP~(\ref{lp:0}). By the last two constraint families, we know that $x^i:=\parentheses*{x_{i,l}}_{0\le l\le m}$ is a distribution for every $1\le i\le n$. From the first two constraint families and Fact~\ref{fact:inducible}, we get that there exists an inducible information structure $\pi$ with respective marginals $x^1,\ldots,x^n$. From the definition of $u$, $V\parentheses*{\pi}$ is exactly the value of the objective function, as desired.

This completes the proof that the optimal sender's value can be computed in polynomial time. Finally, given a solution of the LP corresponding to the marginal distributions over posteriors of an optimal signaling scheme satisfying the conditions of Fact~\ref{fact:inducible}, the constructive proof of Fact~\ref{fact:inducible} from~\cite{brooks2022information} provides a way to build a joint information structure in polynomial time (see also Theorem 16 in~\cite{babichenko2016computational} for an identical argument). Since the sender's utility function is modular, the marginal distributions over posteriors specify the sender's expected utility regardless of their coupling, and thus we are done.
\end{proof}

\section{Proof of Lemma~\ref{lem:lp}}
\label{app:lp}
\begin{proof}[Proof of Lemma~\ref{lem:lp}]
Fix a sender-optimal information structure $\pi$. By Lemma~\ref{lem:grid}, assume w.l.o.g.~that $\supp\parentheses*{\pi}\subseteq \Theta^n$. For $w=0,1$ and $\bar{l}=\parentheses*{l_1,\ldots,l_n}\in L$, choose $x_{\bar{l}}^{w}=\Pr_{\pi}[p_1=\tau_{l_1}\wedge\ldots\wedge p_n=\tau_{l_n} | \omega=w]$. Then the last two families of constraints are clearly satisfied. As for the first family, note that for every $1\le i\le n$ and $q\in \Delta_i$:\footnote{We ignore the edge case $\Pr[p|_{D_i} = q]=0$ in which both sides of the equality are $0$.}

\begin{align*}
    &\sum_{\bar{l}\in L:\; p\parentheses*{\bar{l}}|{D_i}=q}\parentheses*{1-q_i}\mu x_{\bar{l}}^{1}\\
    &= \mu\parentheses*{1-q_i}\Pr[p|_{D_i}=q \mid \omega=1 ]\\
    &=\Pr[\omega=1]\Pr[\omega=0\mid p_i=q_i]\Pr[p|_{D_i}= q \mid \omega=1]\\
    &=\Pr[\omega=0\mid p|_{D_i}= q]\Pr[p|_{D_i}= q \wedge \omega=1]\\
    &=\frac{\Pr[ p|_{D_i}= q \wedge \omega=0]\Pr[p|_{D_i}= q \wedge \omega=1]}{\Pr[p|_{D_i}= q]}\\
    &=\Pr[\omega=0]\Pr[\omega=1 \mid p_i=q_i]\Pr[p|_{D_i} = q \mid \omega=0]\\
    &=\sum_{\bar{l}\in L:\; p\parentheses*{\bar{l}}|{D_i}=q} q_i\parentheses*{1-\mu} x_{\bar{l}}^{0},
\end{align*}

where the first and last transitions follow from union of disjoint events, the second follows from definitions, the third is true because receiver $i$ observes the signals of all the receivers in $D_i$ and his marginal posterior is $p_i$, the fourth is by the conditional probability formula, and the fifth is by symmetry.

Moreover, the coefficient of every $u\parentheses*{S\parentheses*{\bar{l}}}$ in the objective function is, by the law of total probability: $$\mu x_{\bar{l}}^1+\parentheses*{1-\mu} x_{\bar{l}}^0=\Pr_{\pi}[p=p\parentheses*{\bar{l}}],$$ as needed.

Conversely, let $\parentheses*{x_{\bar{l}}^w}$ be a feasible solution to LP~(\ref{lp:1}). Define a signaling scheme $\Sigma$ as follows. When $\omega=w\in\{0,1\}$, draw a joint signal $p\parentheses*{\bar{l}}=\parentheses*{\tau_{l_1},\ldots,\tau_{l_n}}$ from the distribution $\parentheses*{x_{\bar{l}}^{w}}_{\bar{l}\in L}$ s.t.~$\tau_{l_{i}}$ is sent directly to receiver $i$ for $i=1,\ldots,n$ (this is possible due to the last two constraint families). It remains to show that every joint signal $p\parentheses*{\bar{l}}$ induces a joint posterior equal to itself.

Indeed, let $p'$ be the posterior induced by some signal $p\parentheses*{\bar{l}}\in\supp\parentheses*{\Sigma}$. Fix a receiver $1\le i\le n$. Let $q:=p\parentheses*{\bar{l}} |_{D_i}$ be the concatenated signal observed by receiver $i$.  We have to show $p_i'=\tau_{l_{i}}$. Let $M:=\Pr[\omega=1\wedge \sigma_i=q]=\mu\sum_{\bar{l}\in L:\; p\parentheses*{\bar{l}}|{D_i}=q} x_{\bar{l}}^1$. Receiver $i$ has the following unconditional probability of observing a joint signal $\sigma_i=q$:\footnote{We ignore the trivial edge case $\tau_{l_{i}}=0$.}

\begin{align*}
    &\Pr[\sigma_i=q]=\mu \Pr[\sigma_i=q|\omega=1] + \parentheses*{1-\mu}\Pr[\sigma_i=q|\omega=0]\\
    &=\mu\sum_{\bar{l}\in L:\; p\parentheses*{\bar{l}}|{D_i}=q} x_{\bar{l}}^1 + \parentheses*{1-\mu}\sum_{\bar{l}\in L:\; p\parentheses*{\bar{l}}|{D_i}=q} x_{\bar{l}}^0=\parentheses*{1+\frac{1-\tau_{l_{i}}}{\tau_{l_{i}}}}M=\frac{M}{\tau_{l_{i}}}.
\end{align*}

Thus, the marginal posterior of receiver $i$ upon observing $\sigma_i=q$ is:

\begin{align*}
    &\Pr[\omega=1|\sigma_i=q]=\frac{\Pr[\omega=1 \wedge \sigma_i=q]}{\Pr[\sigma_i=q]}=\frac{M}{M/\tau_{l_{i}}}=\tau_{l_{i}},
\end{align*}

which completes the proof.
\end{proof}

\section{Proof of Proposition~\ref{thm:nonforest1}}
\label{app:nonforest}
\begin{proof}[Proof of Proposition~\ref{thm:nonforest1}]
We reduce from the decision version of the minimum $b$-union problem. Namely, given a set of $l\ge 2$ elements $[l]$, $r\ge 2$ sets $B_1,\ldots,B_r\subseteq [l]$, and positive integers $b,k$ s.t.~$r>b \ge 2$ and $l>k\ge 2$, one has to decide whether there are $b$ sets among $B_1,\ldots,B_r$ with union of size at most $k$. This problem is known to be $NP$-hard~\cite{vinterbo2002note}. Given an instance of the decision version of the minimum $b$-union problem, construct a Bayesian persuasion instance with two states, multiple binary-action receivers, and an intricate communication hierarchy as follows:

    \begin{itemize}
        \item The prior is uniform (i.e., $\mu=0.5$).
        \item For each $1\le j\le r$, there is a \emph{set receiver} $s_j$ with corresponding $t_{s_j}=1$.
        \item For each $1\le i\le l$, there is an \emph{element receiver} $e_i$ with $t_{e_i}=0.5$.
        \item The edges in $G$ are $\parentheses*{e_i,s_j}$ s.t.~$i\in B_j$.
        \item The sender's utility $u\parentheses*{S}$ is $0$ if $\absolute*{S}<l-k$, $0.5$ if $l-k\le \absolute*{S}<l+b$, and $1$ if $\absolute*{S}\ge l+b$.
    \end{itemize}

We shall show that:

\begin{itemize}
    \item If there are $b$ sets with union size at most $k$, then $V^{*}=\frac{3}{4}$.
    \item Otherwise, $V^{*}\le\frac{3}{4}-\frac{1}{2 l^2 k^2}$.
\end{itemize}

If there were an FPTAS for our problem, we could have gotten a $\parentheses*{1-\epsilon}$-factor multiplicative approximation of $V^{*}$ for every $\epsilon>0$ in time $\poly\parentheses*{\frac{1}{\epsilon},n=r+l,\absolute*{G}=O\parentheses*{rl}}$. In particular, taking $\epsilon=\frac{1}{6 l^2 k^2}$ would allow us to compute $V^{*}$ up to an additive error of at most $\frac{1}{8 l^2 k^2}$ in $\poly\parentheses*{k,r,l}=\poly\parentheses*{r,l}$ time, and thus to solve the decision version of the minimum $b$-union problem in $\poly\parentheses*{r,l}$ time, a contradiction.

Suppose first that there are indices $1\le j_1<\ldots<j_b$ s.t.~$\absolute*{\cup_{1\le q\le b} B_{j_{q}}}\le k$. Consider the signaling scheme which reveals the true state directly to $s_{j_1},\ldots,s_{j_b}$ and sends totally uninformative signals directly to all other receivers. When $\omega=1$, exactly $l+b$ receivers take action $1$ ($s_{j_1},\ldots,s_{j_b}$ and the element receivers), while when $\omega=0$, at least $l-k$ receivers take action $1$ (the element receivers corresponding to elements of $[l]$ outside of $\cup_{1\le q\le b} B_{j_{q}}$). Therefore, $V^{*}\ge\frac{3}{4}$.

    Assume now that $V^{*}\ge\frac{3}{4}-\frac{1}{2 l^2 k^2}$. We claim that there are $b$ sets among $B_1,\ldots,B_r$ that have union size at most $k$. Suppose, towards contradiction, that it is not true. Note that when $\omega=0$, only element receivers may take action $1$, and there are only $l$ of them. Thus, $$\mathbb{E}\brackets*{u\parentheses*{S}\mid \omega=0}\le \frac{1}{2}.$$ Clearly, we also have $$\mathbb{E}\brackets*{u\parentheses*{S}\mid \omega=1}\le 1.$$ Thus, for every $\pi$ with  $V\parentheses*{\pi}\ge\frac{3}{4}-\frac{1}{2 l^2 k^2}$, we must have both: $$\mathbb{E}_{\pi}\brackets*{u\parentheses*{S}\mid \omega=0}\ge \frac{1}{2}-\frac{1}{l^2 k^2}$$ and: $$\mathbb{E}_{\pi}\brackets*{u\parentheses*{S}\mid \omega=1}\ge 1 - \frac{1}{l^2 k^2}.$$
    
    Since there are only $l$ element receivers, and $u$ takes values in $\{0,0.5,1\}$, we must have that conditional on $\omega=1$, w.p.~$\ge 1-\frac{2}{l^2 k^2}$ there are at least $b$ set receivers that have posterior $1$. Thus, conditional on $\omega=1$, w.p.~$\ge 1-\frac{2}{l^2 k^2}$ at least $k+1$ element receivers have posterior $1$. Therefore, the expected number of element receivers with posterior $1$ conditional on $\omega=1$ is at least $k+1-\frac{2\parentheses*{k+1}}{l^2 k^2}>k+0.5$.
    
    Moreover, since $u$ takes values in $\{0,0.5,1\}$, we deduce that conditional on $\omega=0$, w.p.~$\ge 1-\frac{2}{l^2 k^2}$, at least $l-k$ element receivers must take action $1$. In other words, conditional on $\omega=0$, w.p.~$\ge 1-\frac{2}{l^2 k^2}$ at most $k$ element receivers have posterior $< 1/2$. Hence, the expected number of element receivers with posterior $< 1/2$ conditional on $\omega=0$ is at most $k\parentheses*{1-\frac{2}{l^2 k^2}}+ l \cdot \frac{2}{l^2 k^2}<k+0.5$. Together with the previous paragraph, we deduce from linearity of expectation that there must exist an element receiver $e_i$ s.t.: $$\Pr[p_{e_i}<1/2\mid\omega=0]<\Pr[p_{e_i}=1\mid\omega=1].$$ We claim that this is a contradiction. Indeed, the above inequality would imply by Bayes' law:

\begin{align*}
    &2\Pr[\omega=0\mid p_{e_i}<1/2]\Pr[p_{e_i}<1/2]=\frac{\Pr[\omega=0\mid p_{e_i}<1/2]\Pr[p_{e_i}<1/2]}{\Pr[\omega=0]}<\frac{\Pr[\omega=1\mid p_{e_i}=1]\Pr[p_{e_i}=1]}{\Pr[\omega=1]}\\
    &=2\Pr[p_{e_i}=1].
\end{align*}

Therefore:

\begin{align*}
    &\mathbb{E}\brackets*{1-p_{e_i}\mid p_{e_i}<1/2}\Pr[p_{e_i}<1/2]=\Pr[\omega=0\mid p_{e_i}<1/2]\Pr[p_{e_i}<1/2]<\Pr[p_{e_i}=1].
\end{align*}

We have by Bayes-plausibility $\mathbb{E}\brackets*{1-2p_{e_i}}=0$, and thus:

\begin{align*}
    &\mathbb{E}\brackets*{1-p_{e_i}\mid p_{e_i}<1/2}\Pr[p_{e_i}<1/2]\ge \mathbb{E}\brackets*{1-2p_{e_i}\mid p_{e_i}<1/2}\Pr[p_{e_i}<1/2]\\
    &=\mathbb{E}\brackets*{\max\{1-2p_{e_i},0\}}=\mathbb{E}\brackets*{\max\{1-2p_{e_i},0\}-\parentheses*{1-2p_{e_i}}}=\mathbb{E}\brackets*{\max\{2p_{e_i}-1,0\}}\\
    &\ge\Pr[p_{e_i}=1],
\end{align*}

a contradiction to the previous inequality that completes the proof.
\end{proof}

\section{AI Use Disclosure}
The authors acknowledge that they used GPT 5.6 Sol and GPT 6.0 Astra as helper tools. Importantly:

\begin{itemize}
    \item All the proofs appearing in the manuscript were created, written, and proofread by the authors.
    \item Even in the non-technical parts of the paper, the authors did not copy-paste even a single AI-generated sentence.
\end{itemize}

The authors would like to disclose the following uses of the aforementioned AI tools:

\begin{itemize}
    \item Exploring alternative proof ideas and comparing them to the authors' own ideas, as well as making sure that suspected dead-end directions indeed should be abandoned. Notably, the authors were able to save precious research time when the AI tool confirmed that there was no polynomial-time separation oracle for the LP in Lemma~\ref{lem:lp} in the supermodular case even for $O\parentheses*{1}$-sized tree components, which led the authors to shift their efforts towards proving a negative result. While the lack of a separation oracle indicates that a computational hardness result is plausible, the actual proof of Theorem~\ref{thm:super_hard} was created by the~authors.
     \item Suggesting reformulating the proof of Theorem~\ref{thm:super_hard} in terms of lexicographic reductions, as well as suggesting references to papers on lexicographic reductions, following the authors' idea to use essentially the same approach. These references inspired the authors to simplify their original proof (especially concerning the choice of $M_1$).
    \item Improving the introduction.
    \item Proofreading of all parts of the manuscript.
\end{itemize}

\end{document}